\documentclass[12pt]{article}

\input{epsf}

\usepackage[nosort]{cite}
\usepackage[dvips]{graphicx}
\usepackage{amsmath}
\usepackage{amsfonts}
\usepackage{amssymb}
\usepackage{multirow}
\usepackage{array}

\newcommand{\dslash}{\not\!\partial}
\newcommand{\eps}{\varepsilon}

\begin{document}

\setcounter{page}{0}
\thispagestyle{empty}

%%%%%%%%%%%%%%%%%%%%%%%%%%%%%%%%%%%%%%%%%%%%%%%%%%%%%%%%%%%%%%%%%%%%%%%%%%%%%%%
\begin{flushright}
CERN-PH-TH/2007-233\\
SACLAY-T07/149
\end{flushright}

\vskip 8pt

\begin{center}
{\bf \Large {
Discovering the top partners at the LHC 
\vskip 6pt
using same-sign dilepton final states}}
\end{center}

\vskip 10pt

\begin{center}
{\large Roberto Contino$^{a}$ and G\'eraldine
  Servant $^{a,b}$ }
\end{center}

\vskip 20pt

\begin{center}

\centerline{$^{a}$ {\it CERN Physics Department, Theory Division, CH-1211 Geneva 23, Switzerland}}
\vskip 3pt
\centerline{$^{b}${\it Service de Physique Th\'eorique, CEA Saclay, F91191 Gif--sur--Yvette,
France}}
\vskip .3cm
\centerline{\tt roberto.contino@cern.ch, geraldine.servant@cern.ch}
\end{center}

\vskip 13pt

\begin{abstract}
\vskip 3pt
\noindent
A natural, non-supersymmetric solution to the hierarchy problem generically requires
fermionic partners of the top quark with masses not much heavier than $500\, \text{GeV}$.
We study the pair production and detection at the LHC of the top partners
with electric charge $Q_e=5/3$ ($T_{5/3}$) and $Q_e=-1/3$ ($B$), that are
predicted in models where the Higgs is a pseudo-Goldstone boson. 
The exotic $T_{5/3}$ fermion, in particular, is the distinct prediction of a LR custodial parity invariance
of the electroweak symmetry breaking sector.
Both kinds of new fermions decay to $Wt$, leading to a $t\bar{t}WW$ final state.
We focus on the golden channel with two same-sign leptons, and show
that a discovery could come with less than $100\,\text{pb}^{-1}$ (less than $20\,\text{fb}^{-1}$)
of integrated luminosity  for masses  $M=500\,\text{GeV}$ ($M=1\,\text{TeV}$).
In the case of the $T_{5/3}$,  we present a simple strategy
for its reconstruction in the fully hadronic decay chain.
Although no full mass reconstruction is possible for the $B$, we still find that
the same-sign dilepton channel
offers the best chances of discovery compared to other previous searches that used
final states with one or two opposite-sign leptons, and hence suffered from the large $t\bar t$
background.
Our analysis also directly applies to the search of 4th generation $b^\prime$ quarks.
\end{abstract}

\vskip 13pt
\newpage

%%%%%%%%%%%%%%%%%%%%%%%%%%%%%%%%%%%%%%%%%%%%%%%%%%%%%%%%%%%%%%%%%%%%%%%%%%%%%

\section{Introduction}
\label{sec:introduction}

If one looks at
the formidable legacy left by the LEP experiments, probably the most
precious clue to unravel the mystery on the nature 
of the Electroweak Symmetry Breaking (EWSB) is the evidence, 
although not yet conclusive, in favor of a light Higgs~\cite{LEPEWWG}. 
According to the modern understanding of field theories, combined with an intuitive
naturalness criterion, the existence of a light scalar in the low-energy spectrum,
such as a light Higgs boson, is a clear indication of a highly non-trivial completion
of the Standard Model (SM), with a new symmetry and new particles.
Or, it might be the sign of a dramatic failure of 
naturalness arguments~\cite{Weinberg:1987dv,Agrawal:1997gf,ArkaniHamed:2004fb}.

The most notorious example of symmetry protection for the light Higgs
is Supersymmetry: according to its paradigm, the radiative correction
of each SM field to the Higgs mass is fine tuned against that of a superpartner
of opposite statistics. The top quark contribution, in particular, is balanced by
the contribution of its scalar partners, the stops.
Another kind of symmetry protection, however, could be at work: the light Higgs could be
the pseudo-Goldstone boson of a spontaneously broken global 
symmetry~\cite{Weinberg:1972fn,Georgi:1975tz,Kaplan:1983fs}.
In this case the radiative correction of the top quark to the Higgs mass is balanced 
by the contribution of new partners of the same spin. 
The naturalness criterium suggests that these new heavy fermions 
should have masses below, or not much heavier than, 1 TeV.
It is the production of these top partners at the LHC that we want to study in this paper.

Particularly motivated  is the possibility that
the spontaneous breaking of the global symmetry and the new states originate
from a strongly-coupled dynamics. This would allow for a complete 
resolution of the Hierarchy Problem without the need of fundamental scalar fields,
and would make it possible to generate a large enough quartic coupling for the Higgs
via radiative effects. 
As suggested by the theoretical developments on the connection between gravity 
in higher-dimensional curved spacetimes and strongly-coupled gauge theories~\cite{AdSCFT1,AdSCFT3}, 
the strong dynamics that generates the light Higgs could be 
realized by the bulk of an extra dimension~\cite{Contino:2003ve}.
These extra-dimensional theories are not only fascinating because of the profound
impact they would have on our understanding of high-energy physics, but are also 
extremely interesting as they admit, under certain assumptions,
a perturbative expansion that allows one to compute
several observables of key interest, such as for example the Higgs potential.

The LEP precision data are once again crucial in guiding our theoretical investigation, 
as they seem to be compatible only with a specific kind of strong dynamics: the new
sector must possess a custodial symmetry $G_{C}=$SU(2)$_C$ to avoid large 
tree-level corrections to the $\rho$ parameter~\cite{Sikivie:1980hm}. This in turn implies an unbroken
SU(2)$_L \times$SU(2)$_R\times$U(1)$_X$ invariance of the strong dynamics before EWSB,
meaning that its resonances, in particular the heavy partners of the top quark, will
fill multiplets of such symmetry.
It has been recently pointed out~\cite{Agashe:2006at} that possible modifications to the 
$Z\bar b_Lb_L$ coupling 
can also be substantially suppressed, and the relative LEP constraint more easily satisfied, 
if the custodial symmetry of the strong sector includes a $LR$ parity, $G_{C}=$SU(2)$_C\times P_{LR}$.
More precisely, the $Z\bar b_L b_L$ vertex will not receive zero-momentum corrections from the strong 
dynamics if $b_L$ couples linearly to a composite fermionic operator transforming as a 
$(\mathbf{2},\mathbf{2})_{2/3}$ under SU(2)$_L \times$SU(2)$_R\times$U(1)$_X$ (hypercharge being
defined as $Y=T^3_R+X$).
In this case, as explicitly illustrated by the 5-dimensional models built to incorporate the 
$P_{LR}$ protection~\cite{Contino:2006qr,Carena:2006bn,Panico:2008bx,others}, the heavy partners of 
$(t_L,b_L)$ can themselves fill a $(\mathbf{2},\mathbf{2})_{2/3}$ representation. 
The latter consists of two SU(2)$_L$ doublets: the first, $(T,B)$, has the quantum numbers
of $(t_L,b_L)$; the second -- its ``custodian'' -- is made of one fermion with exotic electric charge 
$Q_e = +5/3$, $T_{5/3}$, and one with charge $Q_e = +2/3$, $T_{2/3}$.
Since the Higgs transforms like a $(\mathbf{2},\mathbf{2})_{0}$, the partners of $t_R$, if any, will form a 
$(\mathbf{1},\mathbf{1})_{2/3}$ or a $[(\mathbf{1},\mathbf{3})\oplus (\mathbf{3},\mathbf{1})]_{2/3}$
of SU(2)$_L \times$SU(2)$_R\times$U(1)$_X$~\cite{Agashe:2006at}.

As explained in detail in Section \ref{section:model}, these new fermions are expected to couple strongly to 
the third generation SM quarks plus one longitudinal $W$, $Z$ gauge boson or the Higgs.
These interactions are responsible for both their single production in hadron collisions and 
their decay, while pair production will proceed via QCD interactions.
The production at the LHC of the heavy fermions with electric charge $+2/3$ 
(the heavy tops $\tilde T$, $T$, $T_{2/3}$)
has been studied in detail in the literature, mainly because of their role in Little Higgs models.
Pair production of the SU(2)$_L$ singlet $\tilde T$, $gg, q\bar q \to \tilde T\bar{\tilde T}$,
was considered in~\cite{AguilarSaavedra}, focussing on final states with one charged lepton.
The process with both heavy tops decaying to $Wb$ was found to be the most promising, though
channels with one neutral decay to $Z$ or $h$ help increase the discovery reach as well.
The minimum integrated luminosity to have a $5\sigma$ statistical significance, 
${\cal S}/\sqrt{\cal B} = 5$, was found to be $L_{min}(5\sigma) = 2.1\, \text{fb}^{-1}$
($90\, \text{fb}^{-1}$) in the case of a heavy top with mass $M_{\tilde T} = 500\, \text{GeV}$
($1\, \text{TeV}$).
As found in Ref.~\cite{Carena:2007tn}, the significance is enhanced if the $\tilde T\bar{\tilde T}$ 
pair-production cross section receives an additional contribution from the exchange of
a heavy gluon.
Single production via $bW$ fusion, $q b\to q' \tilde T$,
was considered in Refs.~\cite{Azuelos:2004dm}, focussing on leptonic final states.
It was found to extend the discovery reach 
to $M_{\tilde T} = 2\, (2.5)\, \text{TeV}$, for $L = 300\, \text{fb}^{-1}$
and a value of the $\tilde TbW$ coupling equal to $\lambda_{\tilde T} = 1\, (2)$.

Pair production of the heavy fermion with electric charge $-1/3$ (the heavy bottom $B$) 
has also been recently considered in~\cite{Dennis:2007tv,Skiba:2007fw}.
\footnote{See also~\cite{ATLAS:TDR2} for an earlier study.}
The process $gg,q\bar q \to B\bar B\to W^- t W^+\bar t$ leads to 
spectacular events with $4W$'s and two bottom quarks, though its observability into
final states with one charged lepton or two leptons with opposite charge is challenged
by the large $t\bar t+jets$ SM background.
To get rid of the latter, Refs.~\cite{Dennis:2007tv} and \cite{Skiba:2007fw} performed
hard cuts on the total effective mass respectively of the jets and of the entire event. 
Ref.~\cite{Skiba:2007fw} also proposed the use of the single-jet invariant mass distribution as a strategy
to further isolate the signal events and reconstruct the hadronically decayed $B$.
The basic idea is that the top and the $W$ originating from the decay of a very massive
$B$ are highly boosted, and the quarks emitted in their hadronic decay will
merge into a single jet with invariant mass $M_j$ close to $m_W$ or $m_t$.

\begin{figure}[t!]
\begin{center}
\includegraphics[width=7.5cm]{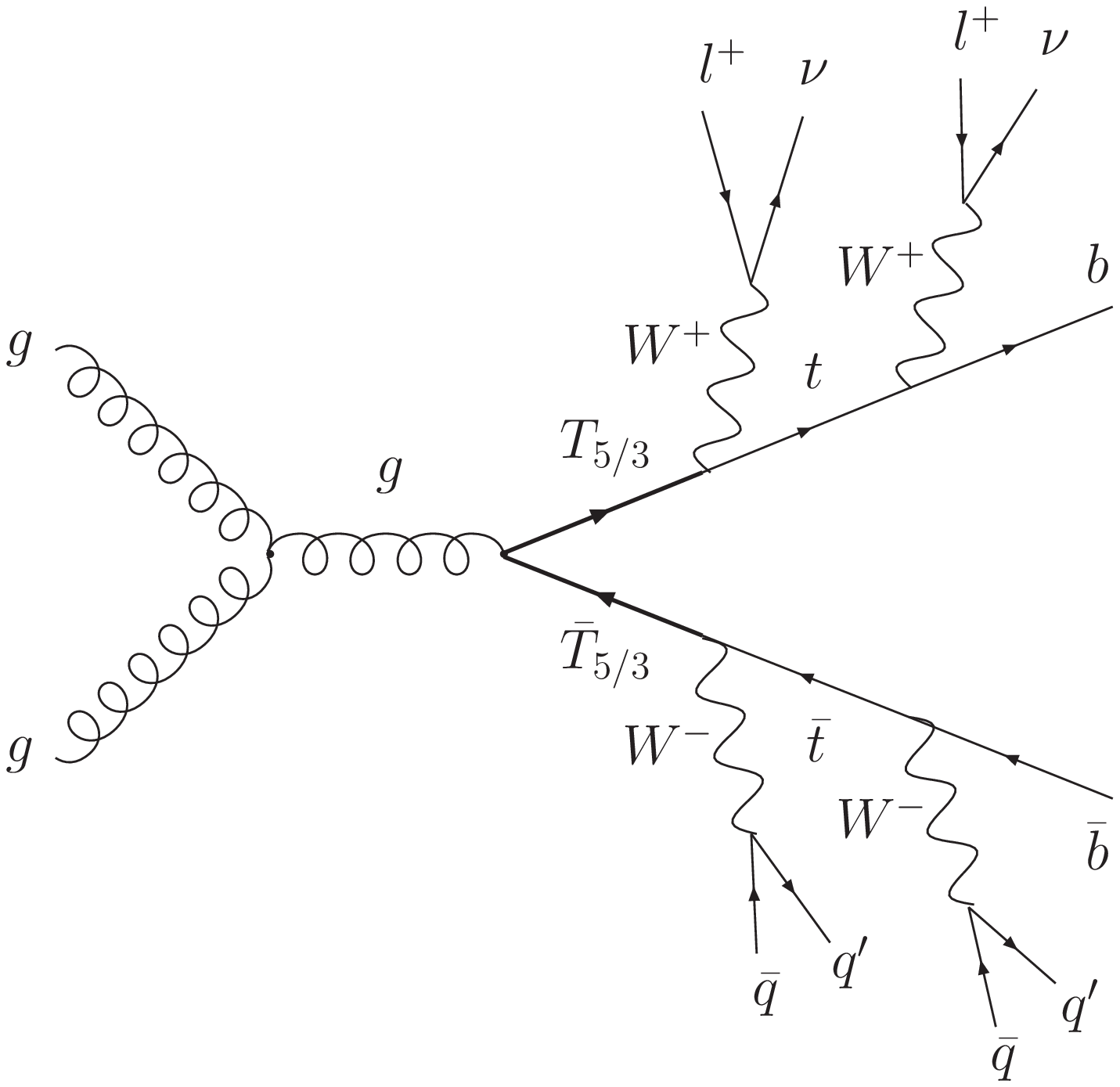}
\hspace{0.5cm}
\includegraphics[width=7.5cm]{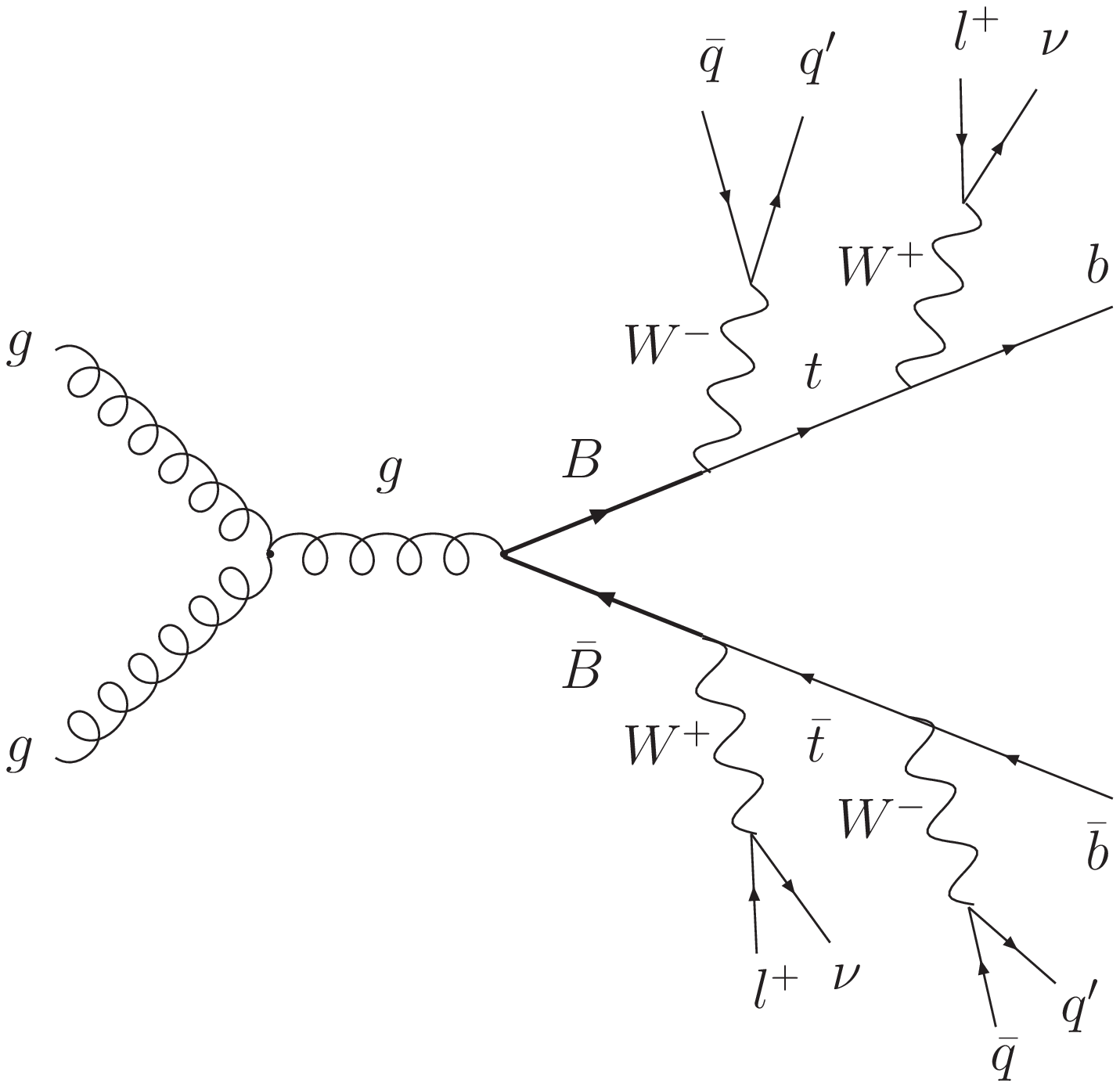}
\caption{\label{fig:diagrams}  \small
Pair production of $T_{5/3}$ and $B$ to same-sign dilepton final states.
}
\end{center}
\end{figure}

In this paper we want to study the pair production of the $B$ and of its custodial partner $T_{5/3}$
proposing a different strategy to get rid of the $t\bar t+jets$ background: looking at final states
with two \textit{same-sign} leptons.
Once pair produced, both the heavy
bottom $B$ and the exotic $T_{5/3}$ decay to $W^+W^+W^-W^-b\bar b$, although with different
spatial configurations as dictated by their different electric charges, see Fig.~\ref{fig:diagrams}.
In the case of the $T_{5/3}$ the two same-sign leptons come from the decay of the same heavy fermion, 
allowing for a full reconstruction of the hadronically-decaying $T_{5/3}$,
while in the case of the heavy bottom they come from different $B$'s.
Despite the fact that a full reconstruction of the $B$ is not possible, we still find that 
the same-sign dilepton channel is probably the most promising one for its discovery.
%%%%%%%%%%%%%%%%%%%%%%%%%%%%%%%%%%%%%%%%%%%%%%%%%%%%%%%%%%%%%%%%%%%%%%%%%%%%
%%%%%%%%%%%%%%%%%%%%%%%%%%%%%%%%%%%%%%%%%%%%%%%%%%%%%%%%%%%%%%%%%%%%%%%%%%%%

In the next section we present a simple effective lagrangian 
for the top partners valid at low energy.
We then describe our Monte Carlo simulation (section \ref{sec:Simulation}),
and define our strategy (section \ref{sec:Strategy}). 
Sections~\ref{subsec:discovery} and~\ref{subsec:reconstruction} present our main analysis: 
first, we show the optimal cuts and characterize the best observables for discovering the heavy 
$T_{5/3}$ and $B$ without making any sophisticated reconstruction;
then, we reconstruct the $W$ and $t$ candidates and pair them to 
reconstruct the $T_{5/3}$ invariant mass. 
We conclude with a critical discussion of our results.

\section{A simple model for the top partners}
\label{section:model}

Although the main results of our analysis will be largely independent of the specific realization
of the new sector, we will adopt as a working example the ``two-site'' description of 
Ref.~\cite{Contino:2006nn}, which reproduces the low-energy regime of the 5D models 
of~\cite{Contino:2006qr,Carena:2006bn}
(see also~\cite{Barbieri:2007bh} for an alternative 4D construction).
Its two building blocks are the weakly-coupled sector of the elementary fields $q_L=(t_L,b_L)$
and $t_R$, and a composite sector comprising two heavy multiplets $(\mathbf{2},\mathbf{2})_{2/3}$, 
$(\mathbf{1},\mathbf{1})_{2/3}$ plus the Higgs (the case with partners of the $t_R$ in a 
$[(\mathbf{1},\mathbf{3})\oplus (\mathbf{3},\mathbf{1})]_{2/3}$ can be similarly worked out):
\begin{equation}
{\cal Q} = (\mathbf{2},\mathbf{2})_{2/3} = 
 \begin{bmatrix} T & T_{5/3} \\ B & T_{2/3} \end{bmatrix} \, , \qquad
\tilde T = (\mathbf{1},\mathbf{1})_{2/3}\, , \qquad 
{\cal H} = (\mathbf{2},\mathbf{2})_{0} = 
 \begin{bmatrix} \phi_0^\dagger & \phi^+ \\ - \phi^- & \phi_0 \end{bmatrix} \, .
\end{equation}
The two sectors are linearly coupled through mass mixing terms, resulting in SM and heavy mass eigenstates
that are admixtures of elementary and composite modes.
The Higgs doublet couples only to the composite fermions, and its Yukawa interactions 
to the SM and heavy eigenstates arise only via their composite component. The Lagrangian in the 
elementary/composite basis is (we omit the Higgs potential and kinetic terms and we assume,
for simplicity, the same Yukawa coupling for both left and right composite chiralities):
\begin{equation} \label{eq:TSlagrangian}
\begin{split}
{\cal L} =& \bar q_L \!\dslash\, q_L + \bar t_R \!\dslash\, t_R \\
 &+ \text{Tr} \left\{ \bar{\cal Q} \left( \dslash - M_Q \right) {\cal Q}  \right\}
 + \bar{\tilde T}  \left( \dslash - M_{\tilde T} \right) \tilde T
 + Y_* \, \text{Tr} \{ \bar{{\cal Q}} \,  {\cal H} \} \,  \tilde{T}  +  h.c \\
 &+ \Delta_{L}\, \bar q_L \left( T,B\right)  + \Delta_{R}\, \bar t_R \tilde T + h.c.
\end{split}
\end{equation}
where  $M_Q$, $M_{\tilde T}$ are the masses of the composite states, $Y_*$ their Yukawa coupling and $\Delta_{L}$, $\Delta_{R}$ are the mixing masses between elementary and composite fields.
After rotating to the mass eigenstate basis, the Yukawa Lagrangian reads
(now denoting with $q_L$, $t_R$ the SM fields, and with $T$, $B$, $T_{5/3}$, $T_{2/3}$,
$\tilde T$ the heavy mass eigenstates):
\begin{equation} \label{eq:yuk}
\begin{split}
{\cal L}_{yuk} = 
& Y_* \sin\varphi_L \sin\varphi_R \left( \bar t_L \phi^\dagger_0 t_R - \bar b_L \phi^- t_R \right) 
 +Y_* \cos\varphi_L \sin\varphi_R \left( \bar T \phi^\dagger_0 t_R - \bar B \phi^- t_R \right) \\
&+Y_* \sin\varphi_L \cos\varphi_R \left( \bar t_L \phi^\dagger_0 \tilde T - \bar b_L \phi^- \tilde T \right)
 +Y_* \sin\varphi_R \left( \bar T_{5/3} \phi^+ t_R + \bar T_{2/3} \phi_0 t_R \right) + \dots
\end{split}
\end{equation}
Here the dots stand for terms with two heavy fermions, and $\sin\varphi_{L,R}$ denote the degree
of compositeness of the SM $t_{L,R}$ quarks: $\tan\varphi_{L} = \Delta_{L}/M_{Q}$,
$\tan\varphi_{R} = \Delta_{R}/M_{\tilde T}$~\cite{Contino:2006nn}.
Equation (\ref{eq:yuk}) explicitly illustrates the specific pattern expected for the couplings of the 
heavy fermions: they couple to one (third-generation) SM quark of defined chirality plus
one longitudinal $W$ or $Z$ boson, or the Higgs.
The values of the couplings are linked to the SM top Yukawa coupling $y_t$; in the two-site
model, in particular, the largest couplings are to the SM fermions with the largest composite
component. For example, if $1 < Y_{*} \ll 4\pi$ -- as one naturally expects if the heavy fermions
are bound states of a strongly coupled sector -- 
the couplings of $T$, $B$, $T_{5/3}$, $T_{2/3}$ are large in the limit of $t_R$ mainly composite, 
$Y_{*}\cos\varphi_L\sin\varphi_R \simeq Y_{*} \sin\varphi_R \gg y_{t}$, while those of $\tilde T$ are 
suppressed~\cite{Contino:2006nn}.
Also, the small ratio between the bottom and top quark masses can be easily explained in this scheme
by assuming that the $b_R$ has a very small composite component. This in turn implies that 
any coupling of $b_R$ to the heavy fermions will be suppressed (for that reason we have
omitted $b_R$ and its own partner(s) from the Lagrangian~(\ref{eq:TSlagrangian})).
Finally, notice that the presence of flavour-changing neutral interactions distinguishes
the heavy partners $T$, $B$ from a fourth generation.

%%%%%%%%%%%%%%%%%%%%%%%%%
\begin{figure}[t!]
\begin{center}
\includegraphics[width=6.5cm]{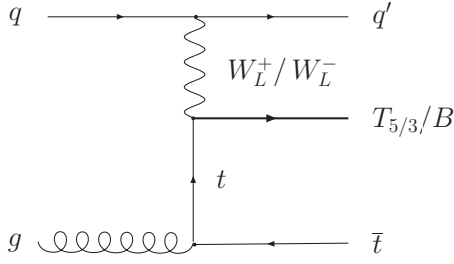}
\caption{\label{fig:singleproduction} \small
Associated single production of $B$ and $T_{5/3}$ at the LHC.}
\end{center}
\end{figure}
%%%%%%%%%%%%%%%%%%%%%%%%%

\begin{figure}[t!]
\begin{center}
\includegraphics[width=11.5cm]{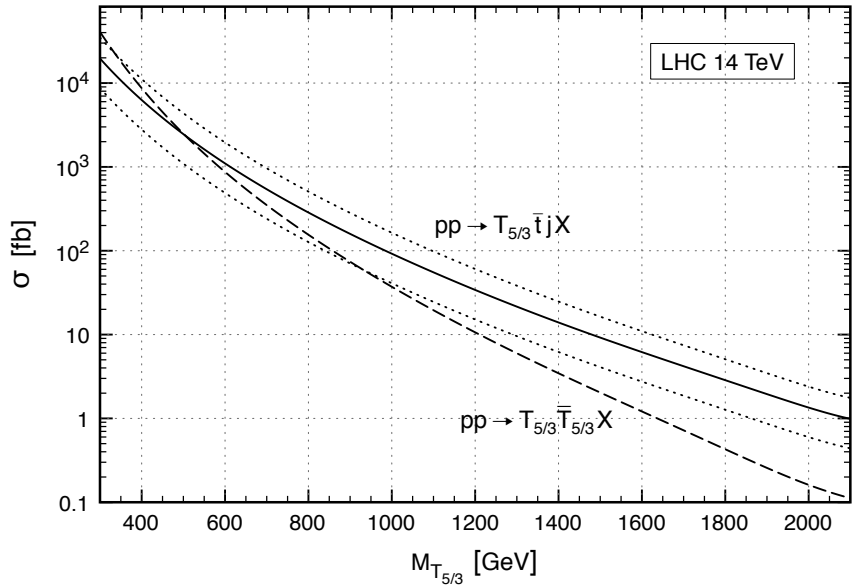}
\caption{\label{fig:crosssection} \small
Production cross sections at the LHC for $T_{5/3}$ as 
functions of its mass. The dashed line refers to pair-production; the solid and the two dotted curves 
refer to single production for the three values of the coupling (from highest to lowest) 
$\lambda_{T_{5/3}}=Y_* \sin \varphi_{R}=4, 3, 2$. 
Cross sections for $B$ are given
by the same curves for the same values of $\lambda_B=Y_* \cos\varphi_{L}\sin \varphi_{R}$.
}
\end{center}
\end{figure}

As anticipated, the interactions of eq.(\ref{eq:yuk}) are responsible for both the decay 
and the single production of the heavy fermions (see for example Ref.~\cite{Contino:2006nn} 
for a more detailed discussion).
Pair production will instead proceed via QCD interactions.
In this work we focus on the pair production of $B$ and $T_{5/3}$ at the LHC, 
considering two values of their mass: $M = 500\,\text{GeV}$ and $M = 1\,\text{TeV}$.
Both $T_{5/3}$ and $B$ decay exclusively to one top plus one longitudinally polarized $W$, with
a decay width
\begin{equation}
\Gamma (T_{5/3} / B\to t_R W_L) = \frac{\lambda^2}{32\pi}\, M\! \left[ \left(1+\frac{m_t^2-m_W^2}{M^2} \right)
 \left(1+\frac{m_t^2 + 2 m_W^2}{M^2} \right) - 4\frac{m_t^2}{M^2} \right] \times \zeta^{1/2} \, , 
\end{equation}
where 
\begin{equation}
\zeta \equiv 1 - 2\, \frac{m_t^2+m_W^2}{M^2} + \frac{\left(m_t^2-m_W^2\right)^2}{M^4}\, ,
\end{equation}
and $M=M_{T_{5/3}}$ ($M=M_B$),
$\lambda = \lambda_{T_{5/3}} = Y_* \sin\varphi_R$ ($\lambda = \lambda_B = Y_* \cos\varphi_L \sin\varphi_R$)
in the case of $T_{5/3}$ ($B$).
For example, setting $\lambda = 3$ gives $\Gamma = 31\, (82)\, \text{GeV}$ for $M=0.5\, (1)\, \text{TeV}$.
Single production proceeds via the diagram of Fig.~\ref{fig:singleproduction},
and becomes dominant for heavier masses, see Fig.~\ref{fig:crosssection}. 
\footnote{Notice that the exact expression for the coupling $\lambda_{T_{5/3}}$ is given by 
$\lambda_{T_{5/3}} = (M_{T_{5/3}}/m_W)  (g/\sqrt{2})  \sin\theta$, where $\sin\theta$ parametrizes the
composite $T_{2/3}$ component of the SM  $t_R$ eigenstate after EWSB, which can be derived by diagonalizing the
$4\times 4$ mass matrix of charge 2/3 fermions. 
Here  we approximate the exact value of $\sin\theta$  at first order in a power series of electroweak
insertions,  which gives $\sin\theta \simeq  v Y_* \sin\varphi_R/\sqrt{2} M_{T_{5/3}}$ for $\sin\theta \ll 1$.
A similar expression can be derived for $\lambda_B$.  This approximation breaks down for
large values of $Y_*$ and light masses $M$, so that $O(1)$ corrections to the value of the single production cross section 
in Fig.~\ref{fig:crosssection} are expected for $M \lesssim 500\,$ GeV and $\lambda\sim 3-4$.
We thank J. A. Aguilaar-Savedra for pointing  this out.
For example, for $M_{T_{5/3}} = 500\,$GeV, $Y_* =4$, $\sin\varphi_R = 0.75$, $\sin\varphi_L = 0.58$, the exact diagonalization
gives $\sin\theta = 0.69$, to be compared with the perturbative value  $(v Y_* \sin\varphi_R/\sqrt{2} M_{T_{5/3}}) = 1.04$. 
Even $O(1)$ corrections to the $T_{5/3}$ and $B$ decay widths, which are   difficult to extract anyway due to 
the finite jet energy resolution, do not affect our analysis.
}
For simplicity, although it is likely to be important for extending the discovery reach to larger masses,
we will neglect single production in the present work. We will argue 
that this should not affect significantly our final results, and that it is in fact a conservative assumption.

Finally, it is worth mentioning that no direct bounds on the heavy quark masses $M_{T_{5/3}}$, $M_B$ exist 
from Tevatron, as no searches have been pursued for new heavy quarks decaying to $tW$. 
The CDF bound on heavy bottom quarks $b'$, $M_{b'}>268\, \text{GeV}$, is derived assuming that $b'$ decays exclusively
to $bZ$~\cite{Aaltonen:2007je}. We estimate that for $M=300\,\text{GeV}$ ($500\,\text{GeV}$), the 
pair-production cross section of $T_{5/3}$ or $B$ at Tevatron is $201\,\text{fb}$ ($1\,\text{fb}$).
For $M=300\,\text{GeV}$ this corresponds to $\sim$ 35 events in the same-sign dilepton channel, before any cut,
with an integrated luminosity of 4 fb$^{-1}$, suggesting that, although challenging, a dedicated analysis at
CDF and D0 could lead to interesting bounds on $M_{T_{5/3}}$, $M_{B}$\footnote{An inclusive search for new physics with same-sign dilepton events was performed recently by CDF using 1 fb$^{-1}$ of data \cite{Abulencia:2007rd}, although these results were not translated into bounds on $B$ and $T_{5/3}$ masses. It will be interesting to see whether the observed slight excess of events persists when using larger data sets.}.

%%%%%%%%%%%%%%%%%%%%%%%%%%%%%%%%%%%%%%%%%%%%%%
%%%%%%%%%%%%%%%%%%%%%%%%%%%%%%%%%%%%%%%%%%%%%%%%%%%%%%%%%%%%%%%%%%%%%%%%%%%%

\enlargethispage{0.5cm}
\section{Signal and Background Simulation}
\label{sec:Simulation}

We want to study the pair production of $B$ and $T_{5/3}$ at the LHC focussing on decay channels with two same-sign 
leptons. We consider two values of the heavy fermion masses, $M=500\,\text{GeV}$ and $M=1\,\text{TeV}$,
and set $\lambda_{T_{5/3}} = \lambda_B = 3$. As explained in the previous section, such large values of the couplings 
are naturally expected if the heavy fermions are bound states of a strongly coupled sector, and $t_R$ is mainly 
composite.
\footnote{For example,
$\lambda_{T_{5/3}} , \lambda_B \simeq 3$ for $Y_* = 3$ and $\sin\varphi_R , \cos\varphi_L \simeq 1$, 
see eq.(\ref{eq:yuk}).}
Notice, however, that our final results will be largely independent of the specific values of $\lambda_{T_{5/3}}$, 
$\lambda_B$, since the latter determine only the decay width of the heavy fermions. 
For our choice of couplings $\Gamma = 31\, (82)\, \text{GeV}$ for $M=0.5\, (1)\, \text{TeV}$.

At the hard-scattering level, the process responsible for pair production to two same-sign leptons is:
\begin{equation}
gg ,\,  q\bar q \to B\bar B ,\, T_{5/3} \bar T_{5/3} \to l^{\pm}\nu\, l^{\pm}\nu\, b\bar b\, q\bar q^\prime q\bar q^\prime \, .
\end{equation}
The physical, observed final state is of the form
\begin{equation} \label{eq:SS2L}
pp\to l^{\pm} l^{\pm} + n\; jets \, + \not\!\! E_T \, , \qquad\quad l = e,\mu \, ,
\end{equation}
where the number of jets depends on the adopted jet algorithm and on its parameters.  
In our analysis we will require $n\ge 5$; this choice will be motivated by 
the distributions and the considerations presented in the next section.
The most important SM backgrounds to the process of eq.(\ref{eq:SS2L}) are $t\bar t W+jets$, $t\bar t WW+jets$
(including the $t\bar t h+jets$ resonant contribution for $m_h\ge 2 m_W$), $WWW+jets$
(including the $W h+jets$ resonant contribution for $m_h\ge 2 m_W$), $W^{\pm}W^{\pm}+jets$
and $Wl^+l^-+jets$ (including the $WZ+jets$ contribution) where one lepton is missed.
To be conservative and consider the case in which the background is largest, we have
set the Higgs mass to $m_h = 180\, \text{GeV}$. This greatly enhances the $t\bar tWW$ and $WWW$ backgrounds.

We have generated both the signal and the SM background events at the partonic level
with MadGraph/MadEvent~\cite{MG-ME},~\footnote{The factorization and renormalization
scales have been chosen as follows: $\mu = M_{T,B} $ for the signal; 
$\mu = 2 m_t+m_W$ for $t\bar tW+jets$; $\mu = 2 m_t+m_h$ for $t\bar tWW+jets$; 
$\mu = m_W + m_h$ for $WWW+jets$; $\mu = 2 m_W$ for $W^{\pm}W^{\pm}+jets$.}
and we have used Pythia~\cite{Sjostrand:2006za} for showering and 
to include the initial and final-state radiation (for simplicity, hadronization and underlying event
have been switched off in Pythia). Jets have been reconstructed using F.~Paige's
\texttt{GETJET} cone algorithm with $E_T^{min}=30\, \text{GeV}$ and two different values 
of the cone size $\Delta R=0.4, 0.7$.
The parton-jet matching has been performed
following the MLM prescription~\cite{MLM}.~\footnote{The full chain of steps in the simulation process 
(linking MadGraph/MadEvent to Pythia, calling of Pythia, jet matching and 
jet reconstruction) has been performed using the
package of dedicated programs in the MadGraph/MadEvent distribution~\cite{MGdistribution}.}
We have not included detector effects in our analysis,
except for a simple gaussian smearing on the jets (we have smeared both the jet energy and momentum absolute
value by $\Delta E/E=100\%/\sqrt{E/\text{GeV}}$, and the jet momentum direction using an angle resolution 
$\Delta \phi=0.05$ radians and $\Delta\eta=0.04$).

The production cross sections for the signal  
and for the various backgrounds are reported in Table~\ref{table:xsecs}.
\begin{table}[t]
\begin{center}
\setlength\extrarowheight{3.5pt}
\begin{tabular}{|p{6.5cm}|c|c|}
\hline
 & $\sigma$ [fb] & $\sigma \times BR(l^\pm l^\pm)$ [fb] \\[0.10cm] \hline 
$T_{5/3}\overline{T}_{5/3} / B {\overline B}+ jets$  \ $(M=500$ GeV)   & $2.5\times 10^3$ & 104 \\[0cm]
$T_{5/3}\overline{T}_{5/3} / B {\overline B}+jets$  \ $(M=1$ TeV)      & 37   & 1.6 \\[0.35cm]
$t {\overline t} W^+W^- + jets$ \ ($\supset t\bar t h + jets$)        & 121  & 5.1 \\[0cm]
$t {\overline t} W^{\pm}+ jets$                                        & 595  & 18.4\\[0cm]
$ W^{+}W^{-} W^{\pm} + jets$ ($\supset h W^{\pm}+jets$)                 & 603  & 18.7 \\[0cm]
$ W^{\pm}W^{\pm} + jets$                                               & 340  & 15.5 \\[0.10cm]
\hline
\end{tabular}
\caption{ \small \label{table:xsecs}
Signal and background cross sections at leading order (left column). The right column
reports the cross section times the branching ratio to two
same-sign leptons final states ($e$ or $\mu$). }
\end{center}
\end{table}
\footnote{
Due to CPU limitations, 
in the case of the $WWW+jets$ and $W^{\pm}W^{\pm}+jets$ backgrounds 
we were not able to generate with MadGraph/MadEvent all the partonic multiplicities required 
for a $5$ jets analysis. In particular, we generated (and matched)
the following partonic processes: $WWW$, $WWWj$, $WWWjj$, and $W^{\pm}W^{\pm}jj$, $W^{\pm}W^{\pm}3j$,
$W^{\pm}W^{\pm}4j$, with $j=$ quark or gluon (notice that the processes $pp\to W^{\pm}W^{\pm}$
and $pp\to W^{\pm}W^{\pm}j$ do not exist due to the conservation of the electric charge).
This means that of the $5$ hard jets required in the analysis, 
one will necessarily originate from Pythia.
This leads to a slight underestimation of these backgrounds, which is however negligible
in our analysis, since $WWW+jets$ and $W^{\pm}W^{\pm}+jets$ are largely subdominant after 
imposing the main cuts of the next section. 
In the case of the leading backgrounds $t\bar tW+jets$ and $t\bar tWW+jets$ all the required 
partonic multiplicities were instead generated. 
}
No K-factors have been included, since those for the backgrounds are not all available
(the K-factor for the signal is $\simeq 1.8\, (1.6)$ for $M=0.5\, (1) \, \text{TeV}$~\cite{Bonciani:1998vc}).
Given its complexity, we were not able to 
fully simulate the $Wl^+l^-+jets$ background,
and for that reason we have not included it in our 
analysis. We have however estimated it as follows.
First, one of the leptons coming from the $l^+l^-$ pair in $Wl^+l^-+jets$ has to be missed in
order for this process to lead to a same-sign dilepton final state. A lepton is considered
missed if it goes outside the electromagnetic calorimeter or the muon chambers ($\eta>2.5$), or if it is
too soft to be detected (see for example~\cite{Maltoni:2002jr}).
In particular, if the lepton is missed because it is soft, it can be arbitrarily close to its companion
in the $l^+l^-$ pair, leading to a logarithmically enhanced collinear contribution if the pair originates
from a virtual photon. 
A naive estimate based on the similar but simpler process $gg,q\bar q\to q\bar q \, l^+l^-$
indicates that, despite this log enhancement in the soft region, the contribution from the photon 
is much smaller than that from the $Z$, after the cuts on the missed lepton are applied.~\footnote{We 
thank Mauro Moretti for poiting this out and for correcting an error in our previous estimate.}
From the $WZ+jets$ cross section we thus expect  
the $Wl^+l^-+jets$ background to be smaller than~$\sim 10-20\%$~of the sum of the
other backgrounds. This includes a $\sim 10\%$ efficiency due to the lepton veto.
While this estimate shows that $Wl^+l^-+jets$ is not entirely negligible, the error due to 
its exclusion is within the uncertainty of our leading-order analysis.
Moreover, the $Wl^+l^-+jets$ cross section is expected to be strongly suppressed after
requiring the reconstruction of one $W$ and one top as done in section~\ref{subsec:reconstruction}. 

%
%
% In particular, if the lepton is missed because it is soft, it can be arbitrarily close to its companion
% in the $l^+l^-$ pair. 
% If the pair originates from a virtual photon, this leads to a logarithmically divergent 
% contribution to the cross section that is cutoff by the finite lepton mass. 
% This argument shows that diagrams with a virtual photon can give a contribution
% to the cross section larger than that of diagrams where the lepton pair originates from a (possibly on-shell) $Z$. 
% We have estimated their relative importance in a similar but simpler process: $gg,q\bar q\to q\bar q \, l^+l^-$.
% We found that in the case of electrons (muons), including the virtual photon exchange enhances the cross section 
% by a factor $\sim 5\, (2)$ compared to the case in which only the $Z$ contribution is retained.
% We were then able to derive a rough estimate of the total $Wl^+l^-+jets$ cross section 
% starting from $WZ+jets$ (which is simpler to simulate): we find that after the main cuts of the next section
% the $Wl^+l^-+jets$ background is expected to be smaller than $\sim 30\%$ of the sum of the
% other backgrounds. This includes a $\sim 10\%$ efficiency due to the lepton veto.
% While this estimate shows that $Wl^+l^-+jets$ is not entirely negligible, the error due to 
% its exclusion is within the uncertainty of our leading-order analysis.
% Moreover, the $Wl^+l^-+jets$ cross section is expected to be strongly suppressed after
% requiring the reconstruction of one $W$ and one top as done in section~\ref{subsec:reconstruction}. 

Another potential source of background are $t\bar t + jets$ events where the charge of one of the two
leptons from the top decays is misidentified. 
Given the large $t\bar t+jets$ cross section, even a charge misidentification probability
$\eps_{mis}\sim \text{a few}\times 10^{-3}$ would result into a same-sign dilepton background of the same
order of $t\bar t W+jets$.~\footnote{Requiring the reconstruction of one $W$ and one top
as in section~\ref{subsec:reconstruction} is however expected to reduce significantly 
more the $t\bar t + jets$ events background than $t\bar t W+jets$ or $t\bar t WW+jets$.}
The value of $\eps_{mis}$ strongly depends on the $p_T$ and on the pseudo-rapidity of the lepton,
and it is typically smaller for muons than for electrons. We find that the hardest lepton in the
$t\bar t+jets$ events has a $p_T$ distribution peaked at values smaller than $100\, \text{GeV}$
(the second hardest lepton has instead a significantly softer $p_T$ distribution).
For such low-$p_T$ leptons we have not found accurate estimates of $\eps_{mis}$
in the literature (studies on charge misidentification usually focus on leptons with very large $p_T$,
from several hundred GeV to a few TeV, for which $\eps_{mis}$ is larger).
From the latest ATLAS and CMS TDRs, probabilities as low as $\sim 10^{-4}$ seem to be realistic in the case
of muons, while slightly larger values are expected for electrons~\cite{CMS:TDR1,ATLAS:TDR1}.
If $\eps_{mis}=10^{-4}$, the $t\bar t+jets$ background would be smaller by one order of magnitude than the dominant $t\overline{t}W+jets$ background in Table \ref{table:xsecs}, hence safely negligible. 
In absence of a realistic estimate of $\eps_{mis}$ as a function of the lepton's $p_T$ and pseudo-rapidity,
we decided not to include the $t\bar t+jets$ background events in our analysis.
It is however clear that a specific and accurate estimate of this background is required to validate 
our results.

Finally, it is worth commenting on the possible background due to possible additional leptons coming from
$b$ decays (that were not taken into account in our analysis).
These leptons have a very soft $p_T$ spectrum, so that the cuts imposed in section~\ref{subsec:discovery}
on the lepton transverse momentum (we require $p_T>25\, \text{GeV}$ for the softest lepton), 
together with the isolation cut $\Delta R_{lj}>0.4$ between 
any lepton and jet, are expected to reduce such background to safely negligible levels.

\section{Defining our Strategy}
\label{sec:Strategy}

\begin{figure}[htb!]
\begin{center}
\includegraphics[width=8.3cm]{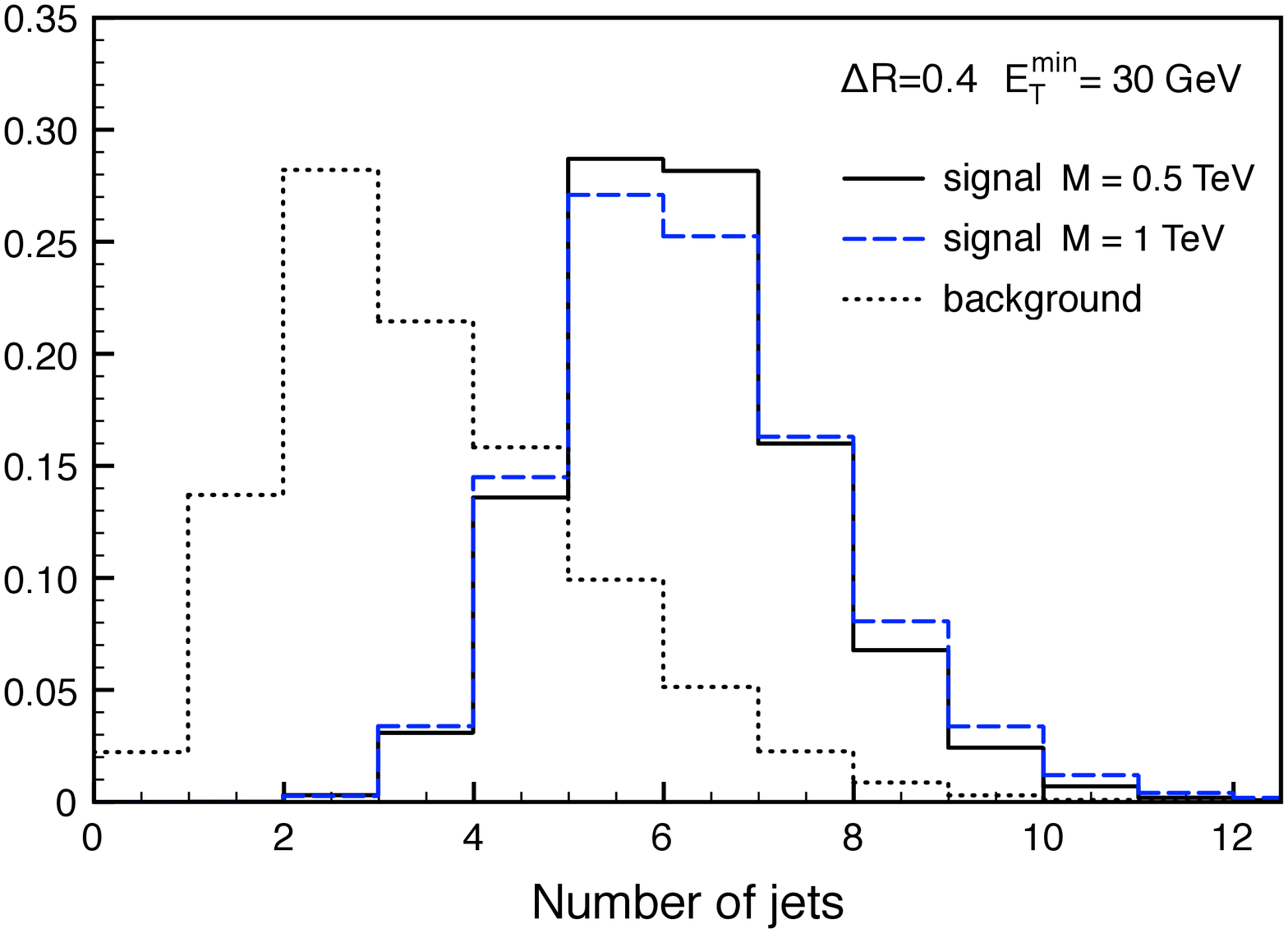} \hspace{-0.35cm}
\includegraphics[width=8.3cm]{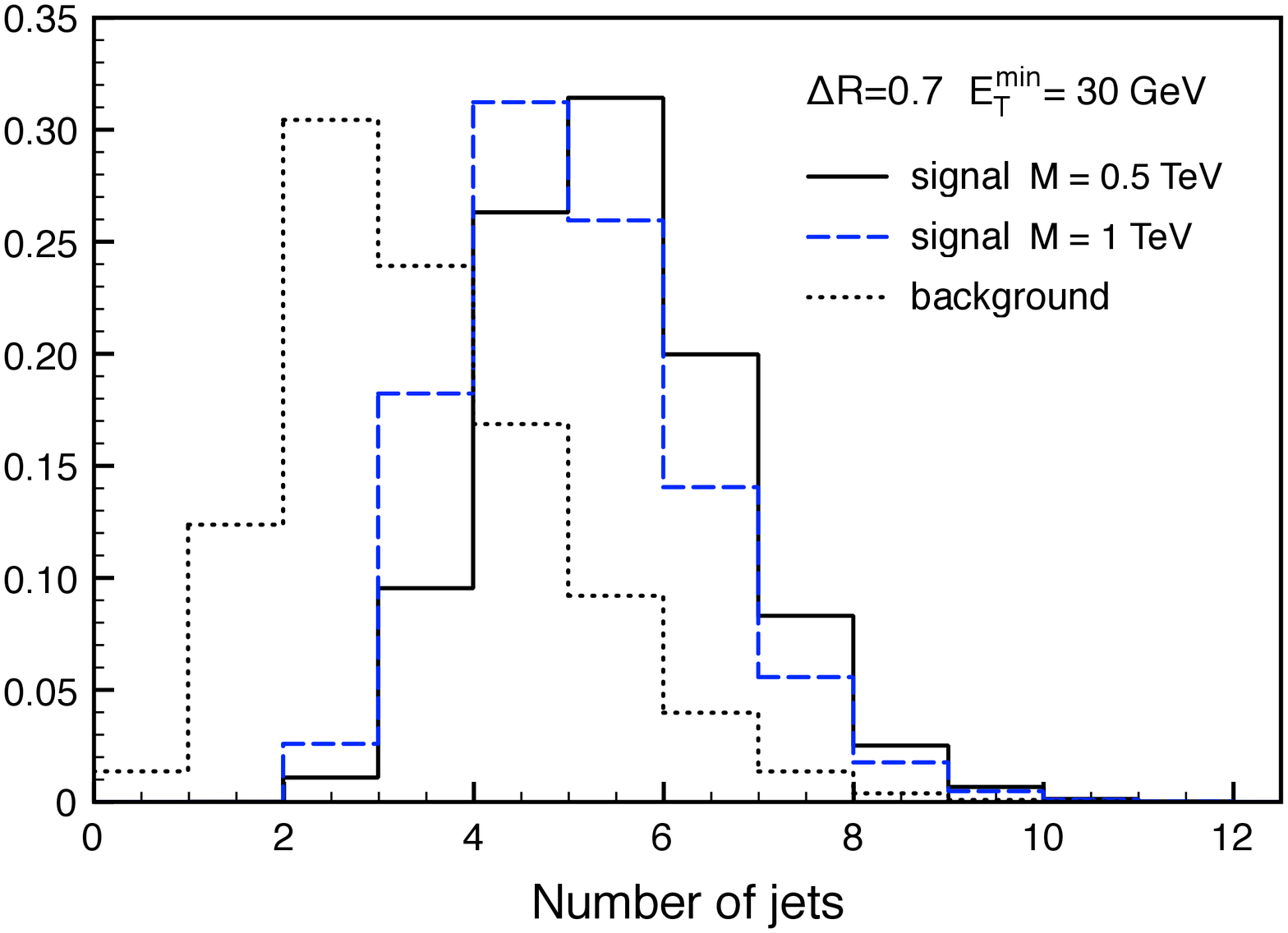}
\caption{\label{fig:Number_jets}  \small
Fractions of signal and background events with a given number of jets 
for $E^{min}_T=30\,\text{GeV}$ and two different jet cone sizes: 
$\Delta R= 0.4$ (left plot), and $\Delta R= 0.7$ (right plot).}
\end{center}
\end{figure}

\begin{figure}[t!]
\begin{center}
\includegraphics[width=8.25cm]{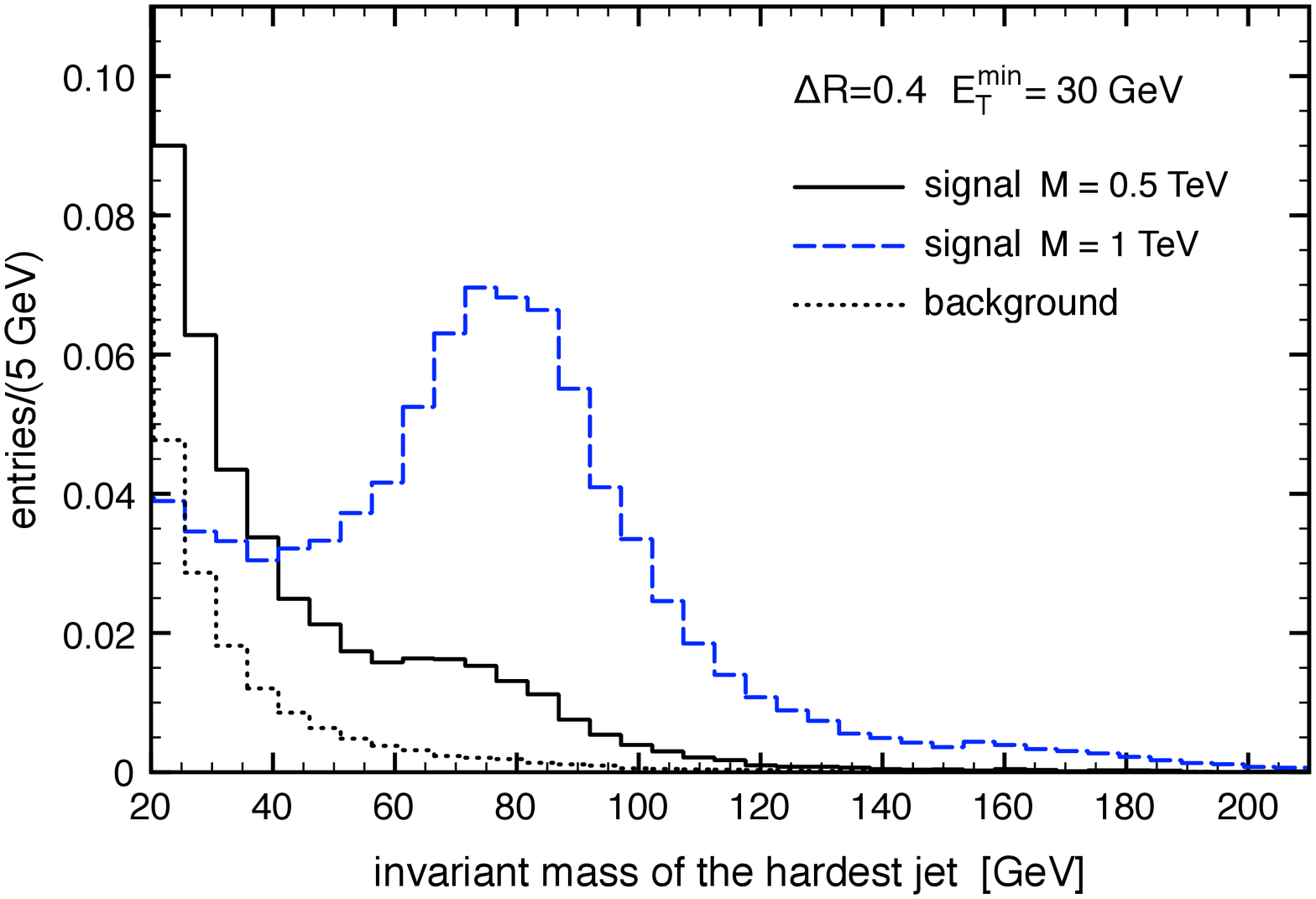} \hspace{-0.3cm}
\includegraphics[width=8.25cm]{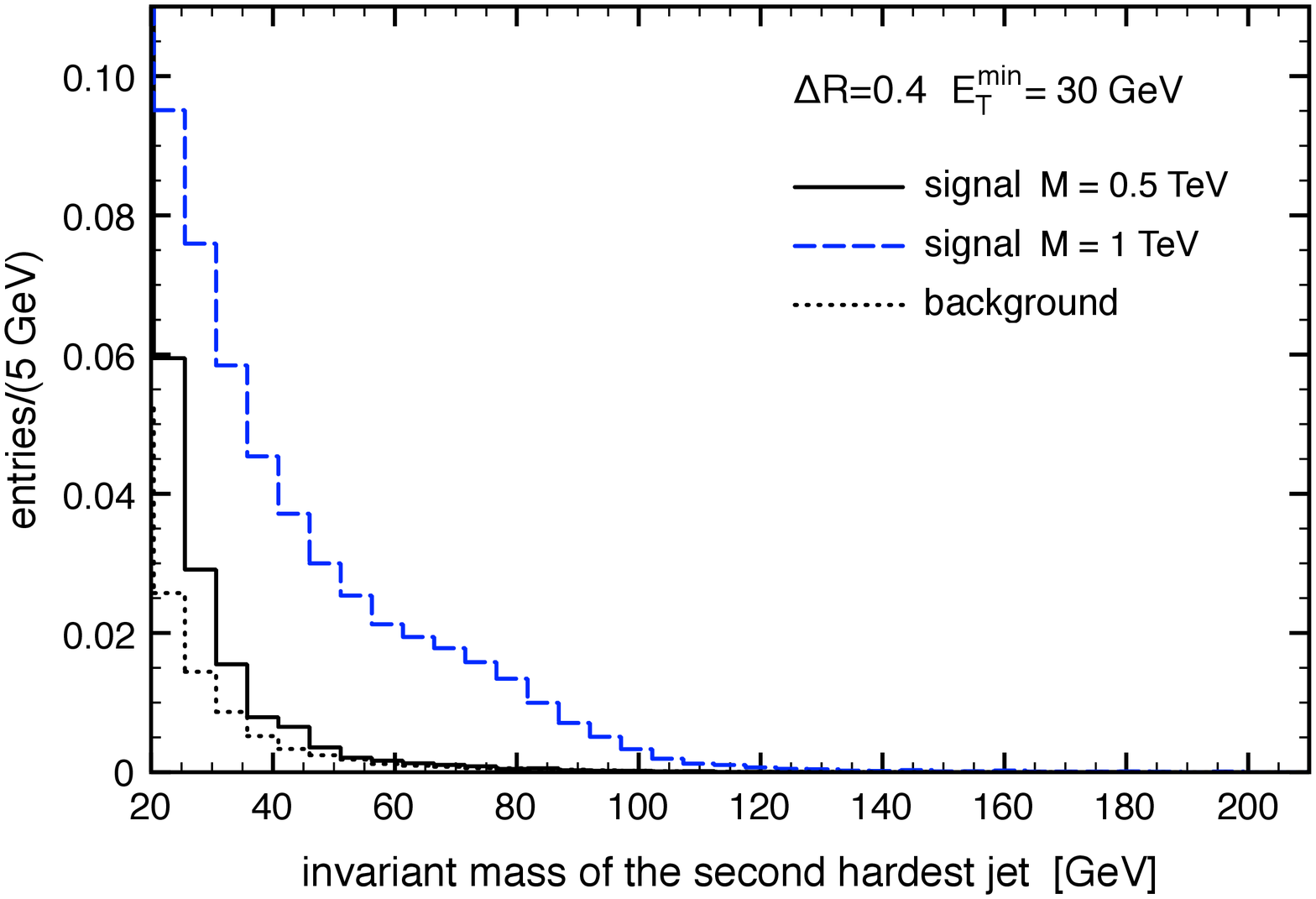} 
\\[0.2cm]
\includegraphics[width=8.25cm]{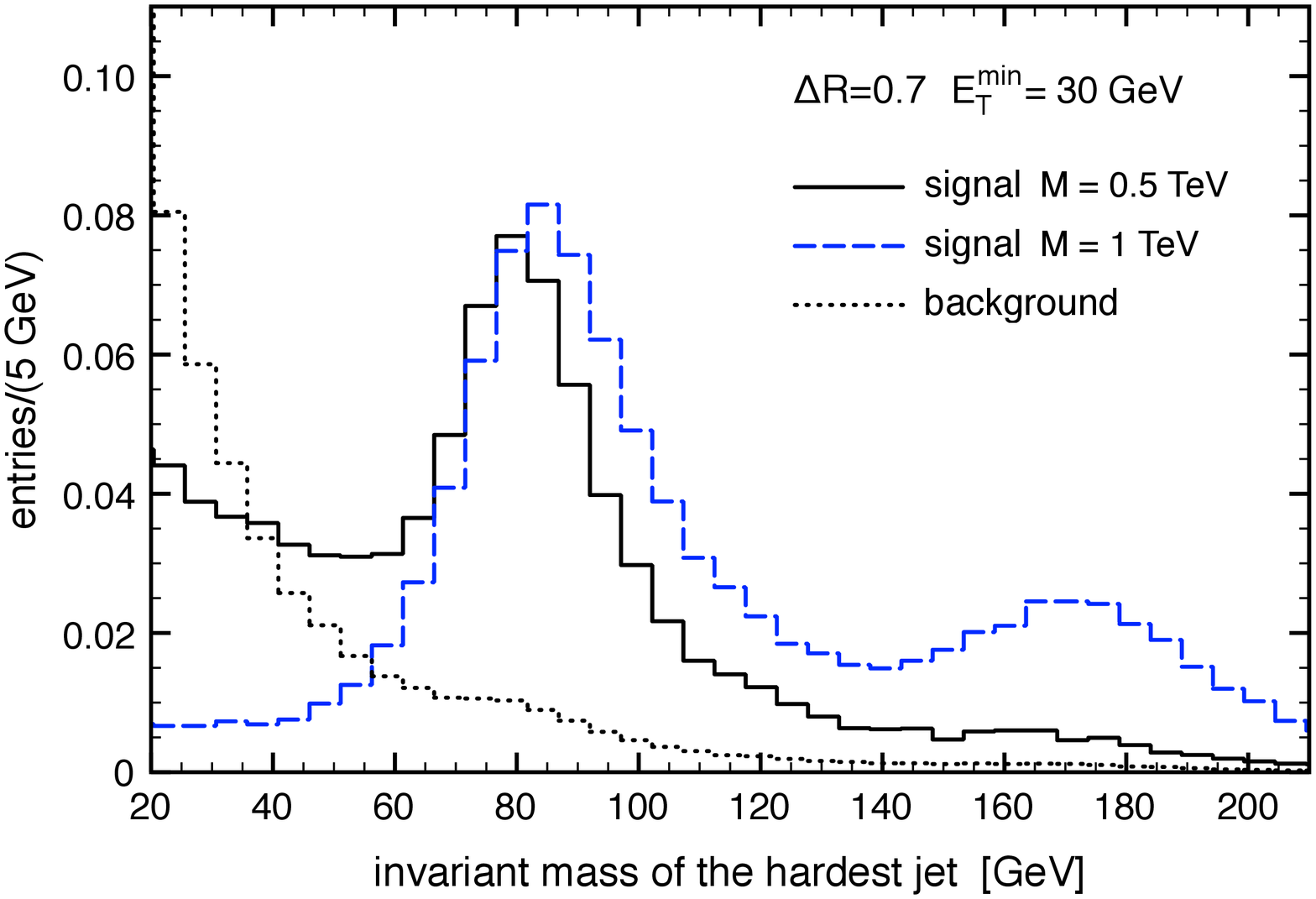} \hspace{-0.3cm}
\includegraphics[width=8.25cm]{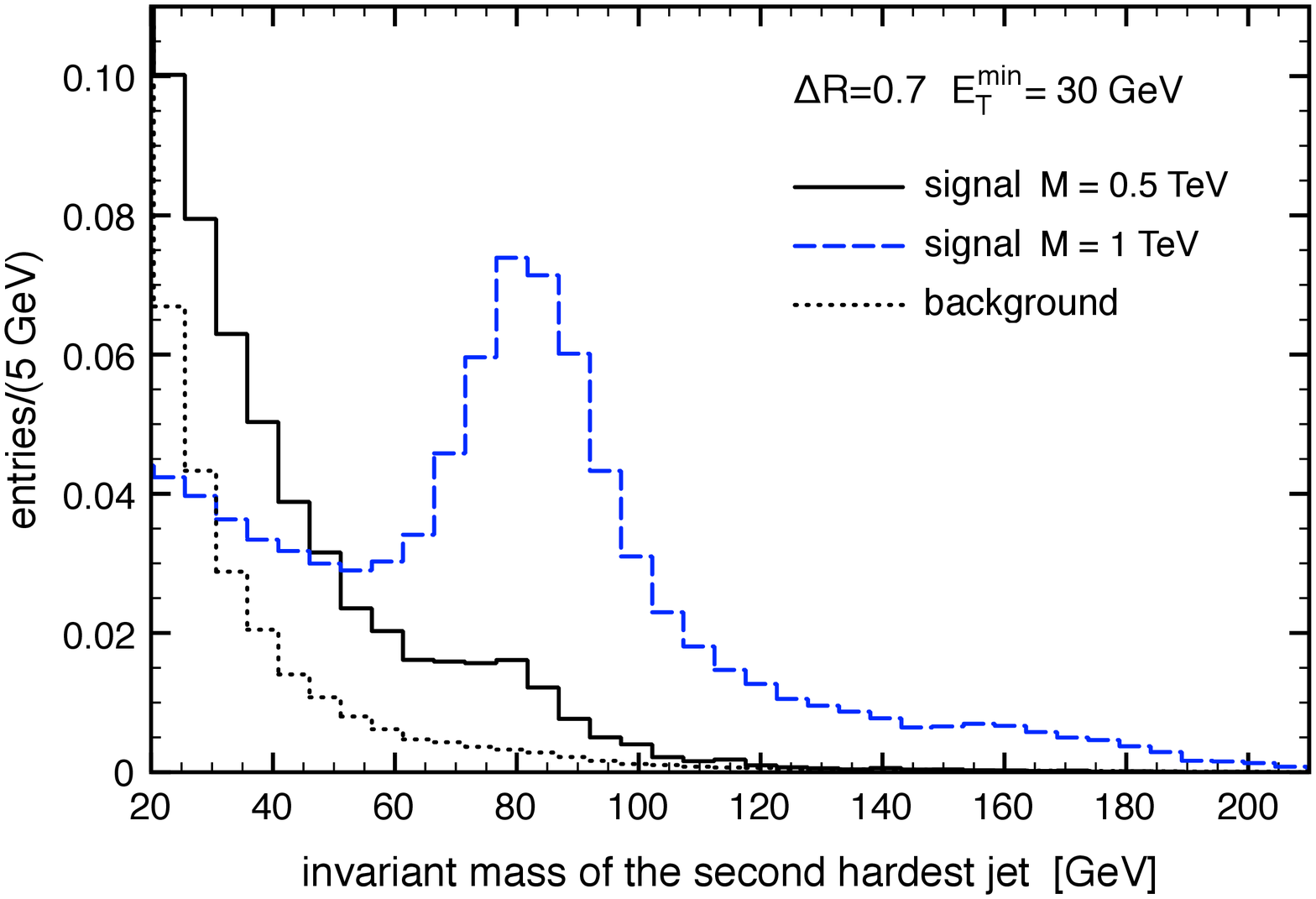}
\caption{\label{fig:InvMassHardestJets} \small
Invariant mass of the first (left) and second (right) hardest jet
for $\Delta R=0.4$ (top) and $\Delta R=0.7$ (bottom). 
All distributions are normalized to unit area.}
\end{center}
\end{figure}

In this section we illustrate our strategy for the analysis of same-sign dilepton events. 
A first important information on the kinematics of signal and background events comes 
from the number of reconstructed jets, showed in Fig.~\ref{fig:Number_jets}
for two different choices of the cone size: $\Delta R=0.4,0.7$.
%%%%%%%%%%%%%%%%%%%%%%%%%%%%%%%%%%%%%%%%%%%%%%%%%%%%%
%%%%%%%%%%%%%%%%%%%%%%%%%%%%%%%%%%%%%%%%%%%%%%%%%%%%%
For $\Delta R=0.4$, the largest fraction of signal events have $5$ or $6$ jets, both in the case of
$M=500\,\text{GeV}$ and of $M=1\,\text{TeV}$ (by signal here we mean either $T_{5/3}\bar T_{5/3}$ or
$B\bar B$ events), while the total background distribution is peaked at smaller values (this is mainly
due to the low jet multiplicity in the $WWW+jets$ and $W^{\pm}W^{\pm}+jets$ backgrounds).
In the case of the signal, the hard scattering process produces $6$ quarks, after the
decay of the top and of the $W$. It turns out that for $M=500\,\text{GeV}$
the 5-jet bin is mostly populated by events where the 6th jet is lost because it is too soft
(i.e. it does not meet the minimum transverse energy requirement, $E_T\ge 30\,\text{GeV}$),
whereas for $M=1\,\text{TeV}$ the 5-jet bin mainly contains
events in which two jets coming from a boosted $W$ decay have merged into a single jet.
This is clearly illustrated in Fig.~\ref{fig:InvMassHardestJets}, which shows the invariant mass spectrum 
of the hardest and next-to-hardest jet.
%%%%%%%%%%%%%%%%%%%%%%%%%%%%%%%%%%%%%%%%%%%%%%%%%%%%%%%%%%%%%%%%%%%%%%%%%%%%
%%%%%%%%%%%%%%%%%%%%%%%%%%%%%%%%%%%%%%%%%%%%%%%%%%%%%%%%%%%%%%%%%%%%%%%%%%%%
For $M=1\,\text{TeV}$ the invariant mass of the hardest jet has a peak in correspondence to $m_W$, while
such a peak is absent in the case of $M=500\,\text{GeV}$, as well as in the distributions of the second
hardest jet. We thus conclude that for a cone size $\Delta R=0.4$, signal events with $M=1\,\text{TeV}$
have one ``double'' jet, while background and $M=500\,\text{GeV}$ signal events have none.

If one enlarges the cone size to $\Delta R=0.7$, the most populated bins in the 
distributions of Fig.~\ref{fig:Number_jets} (right plot)
are those with $4$ and $5$ reconstructed jets.
This suggests that in this case, both for $M=1\,\text{TeV}$ and $M=500\,\text{GeV}$,
another pair of closeby jets, originating from the decay of the second hadronic $W$ decay,
is merged into a single, double jet. 
This is again clearly illustrated in Fig.~\ref{fig:InvMassHardestJets} (bottom plots),
where one can see that signal events with $M=500\,\text{GeV}$ have one double jet with $M_j\simeq m_W$, while
those with $M=1\,\text{TeV}$ have two. There are even cases, for $M=1\,\text{TeV}$, where all three jets 
from the hadronic decay of the top merge into a single jet with $M_j\simeq m_t$, see the left bottom plot
of Fig.~\ref{fig:InvMassHardestJets}. We find, however, that the fraction of these ``triple'' top jets
is relatively small.

Identifying and selecting events with one triple and one double jet was the strategy adopted in 
Ref.~\cite{Skiba:2007fw} to discover and reconstruct heavy bottoms with $M=1\,\text{TeV}$ (see also 
Refs.~\cite{Butterworth:2002tt,Holdom:2007nw,Butterworth:2007ke,Baur:2007ck,Agashe:2007ki,Carena:2007tn} 
where a similar jet-mass technique was applied to different processes). 
According to the authors of Ref.~\cite{Skiba:2007fw}, the presence of these massive jets can 
discriminate the events of the signal from those of the background. In particular, 
an excess of more than $5\sigma$ compared to the SM prediction
can be obtained with $L=100\,\text{fb}^{-1}$ by counting the number of jets with $M_j\sim m_W$. 
This evidence alone, however, is not per se an indication that a heavy $B$ has been
produced (the boosted $W$ could arise from a different process), and it has to be
accompanied by a full reconstruction of the $B$ invariant mass.~\footnote{
The invariant mass distribution of the $Wt$ system is presented in
Ref.~\cite{Skiba:2007fw} only for final states with one lepton, without however 
quoting the statistical significance of the resonant peak.}

The effective validity of such a single-jet mass technique seems to
depend significantly on the adopted jet algorithm and the value of its parameters.
The  $k_T$ algorithm~\cite{Catani:1993hr} was chosen in Ref.~\cite{Skiba:2007fw}.
As the plots of Fig.~\ref{fig:InvMassHardestJets} show, at least one double or triple jet can be resolved 
into individual jets by using a cone algorithm with $\Delta R=0.4$. 
Resolving as many jets as possible seems to be a better option than choosing a larger cone size and
imposing only jet-mass constraints: First, because in the former case the
QCD background will have a larger jet multiplicity (one or two jets more), hence a smaller cross section
to begin with.
Second, because the requirement of having two closeby jets (in our specific case with $\Delta R_{jj} < 0.7$) 
with an invariant mass $M_j\simeq m_w$ (or $M_j\simeq m_t$) should be 
as effective as -- if not more effective than -- 
the cut on the invariant mass of the corresponding double (or triple) jet.
Clearly, measuring the single-jet mass, as well as 
analyzing the jet substructure~\cite{Butterworth:2002tt,Butterworth:2007ke},
\footnote{See also~\cite{Lillie:2007yh,topLH} for the related issue of the identification 
of highly boosted tops.}
remain promising strategies for the cases in which the double jet cannot be resolved.

For the reasons explained above,
we decided to perform our analysis setting the cone size to $\Delta R =0.4$, and to require
at least $5$ reconstructed jets in the final state, both for $M=500\,\text{GeV}$ and $M=1\,\text{TeV}$.
Even in the case of $M=1\,\text{TeV}$, where the signal has typically one double jet,
we preferred not to impose any cut on the single-jet invariant mass,
trying to develop a strategy as independent as possible
of the details of the detector and of the jet algorithm.
In this sense our results are somehow conservative, as one can hope to eventually improve on them
by making use of jet mass cuts.

\section{Discovery Analysis}
\label{subsec:discovery}

In this section and in the next one we present our main analysis of same-sign dilepton events.
We focus first on the discovery of the top partners, proposing a simple strategy that does not
rely on any sophisticated reconstruction, nor does it require $b$-tagging. 
We will adopt $L=10\,\text{fb}^{-1}$ ($L=100\,\text{fb}^{-1}$) as a reference integrated luminosity for 
the various plots in the case $M=500\,\text{GeV}$ ($M=1\,\text{TeV}$), and we will consider two
different scenarios (or ``models''): in the first, both $B$ and the exotic $T_{5/3}$ are present
with the same mass $M$; in the second, only $B$ exists.~\footnote{The case where $T_{5/3}$ is much 
lighter than $B$ is disfavored by electroweak precision data \cite{Barbieri:2007bh}.}

Our main cuts to isolate the signal are the followings:
\begin{equation} \label{maincuts}
\begin{gathered}
\text{\underline{2 same-sign}} \\
\text{\underline{leptons}} \\
\text{\underline{($e$ or $\mu$)}}
\end{gathered} :
\begin{cases}
p_T(\text{1st})  \geq 50 \; \text{GeV} \\
p_T(\text{2nd})  \geq 25 \; \text{GeV} \\
|\eta_l| \leq 2.4 \, ,  \ \  \Delta R_{lj} \geq 0.4
\end{cases} \quad
\text{\underline{jets}}: 
\begin{cases}
p_T(\text{1st})  \geq 100 \; \text{GeV} \\
p_T(\text{2nd})  \geq 80 \; \text{GeV} \\
n_{jet} \geq 5 ,  \ \ |\eta_j| \leq 5
\end{cases} \quad
\not\!\! E_T  \geq 20 \; \text{GeV} \, ,
\end{equation}
where 1st and 2nd refer respectively to the first 
and second hardest jet or lepton (electron or muon).
The relative efficiencies are reported in Table~\ref{table:epsmain}.
%%%%%%%%%%%%%%%%%%%%%%%%%%%%%%%%%%%
\begin{table}[t]
\begin{center}
\setlength\extrarowheight{3pt}
\begin{tabular}{|l|c|c|c|c|c|c|}
\hline
& signal  & signal & \multirow{2}{*}{$t\bar t W$}
& \multirow{2}{*}{$t\bar t WW$} & \multirow{2}{*}{$WWW$} & \multirow{2}{*}{$W^{\pm}W^\pm$} \\
& ($M=500$ GeV) & ($M=1$ TeV) & & & & \\[0.05cm]
\hline 
Efficiencies ($\eps_{main}$) & 0.42 & 0.43 & 0.074 & 0.12 & 0.008 & 0.01 \\[0.01cm]
$\sigma\, \text{[fb]} \times BR \times \eps_{main}$ & 44.2 & 0.67 & 1.4 & 0.62 & 0.15 & 0.16 \\[0.05cm]
\hline 
\end{tabular}
\caption{\label{table:epsmain} \small
Efficiencies of the main cuts of eq.(\ref{maincuts}). Here signal means either
$T_{5/3}\bar T_{5/3}$ or $B\bar B$ events. 
}
\end{center}
\end{table}
%%%%%%%%%%%%%%%%%%%%%%%%%%%%%%%%%%%

\begin{figure}[t]
\begin{center}
\includegraphics[width=8.3cm]{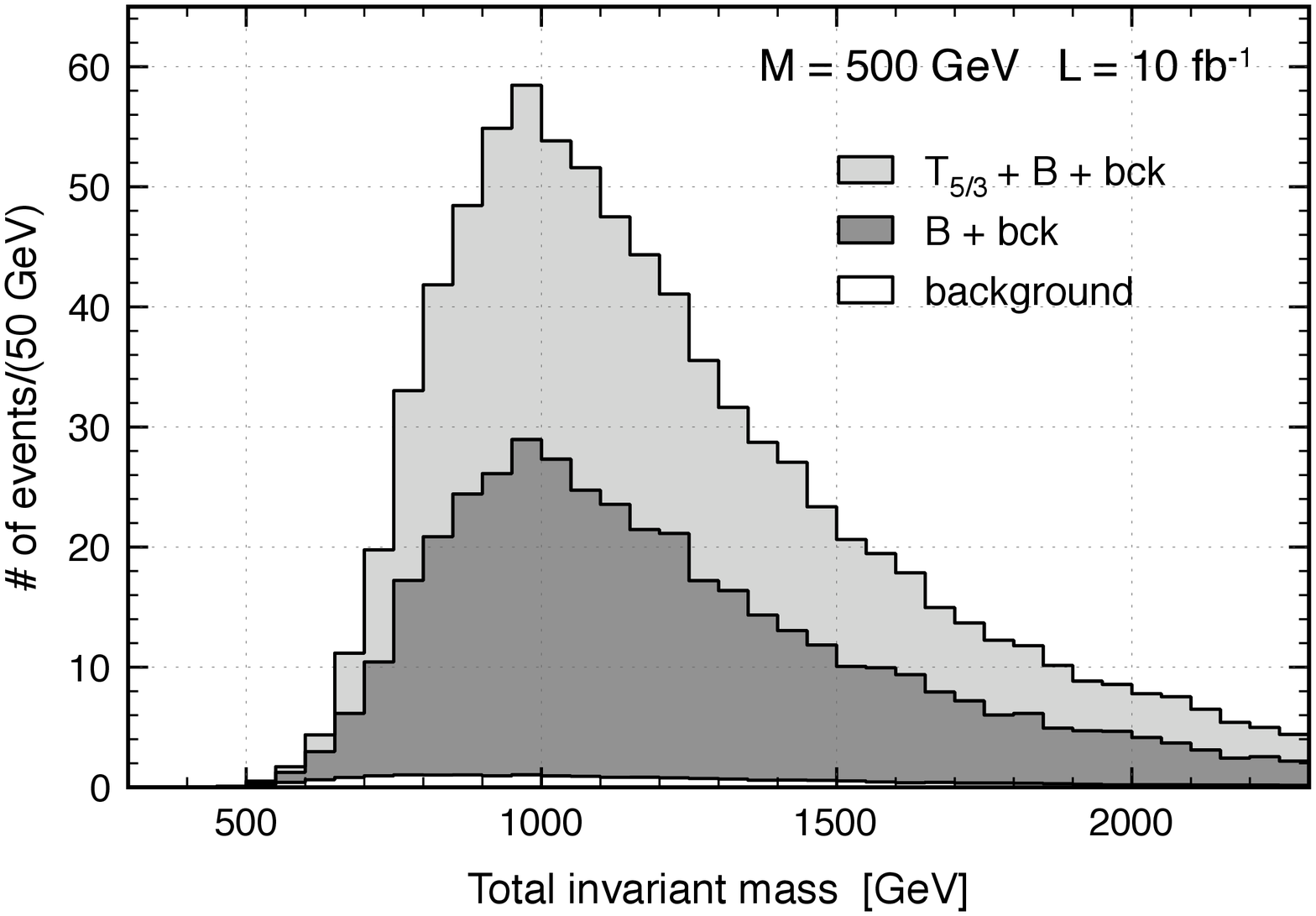} \hspace{-0.35cm}
\includegraphics[width=8.3cm]{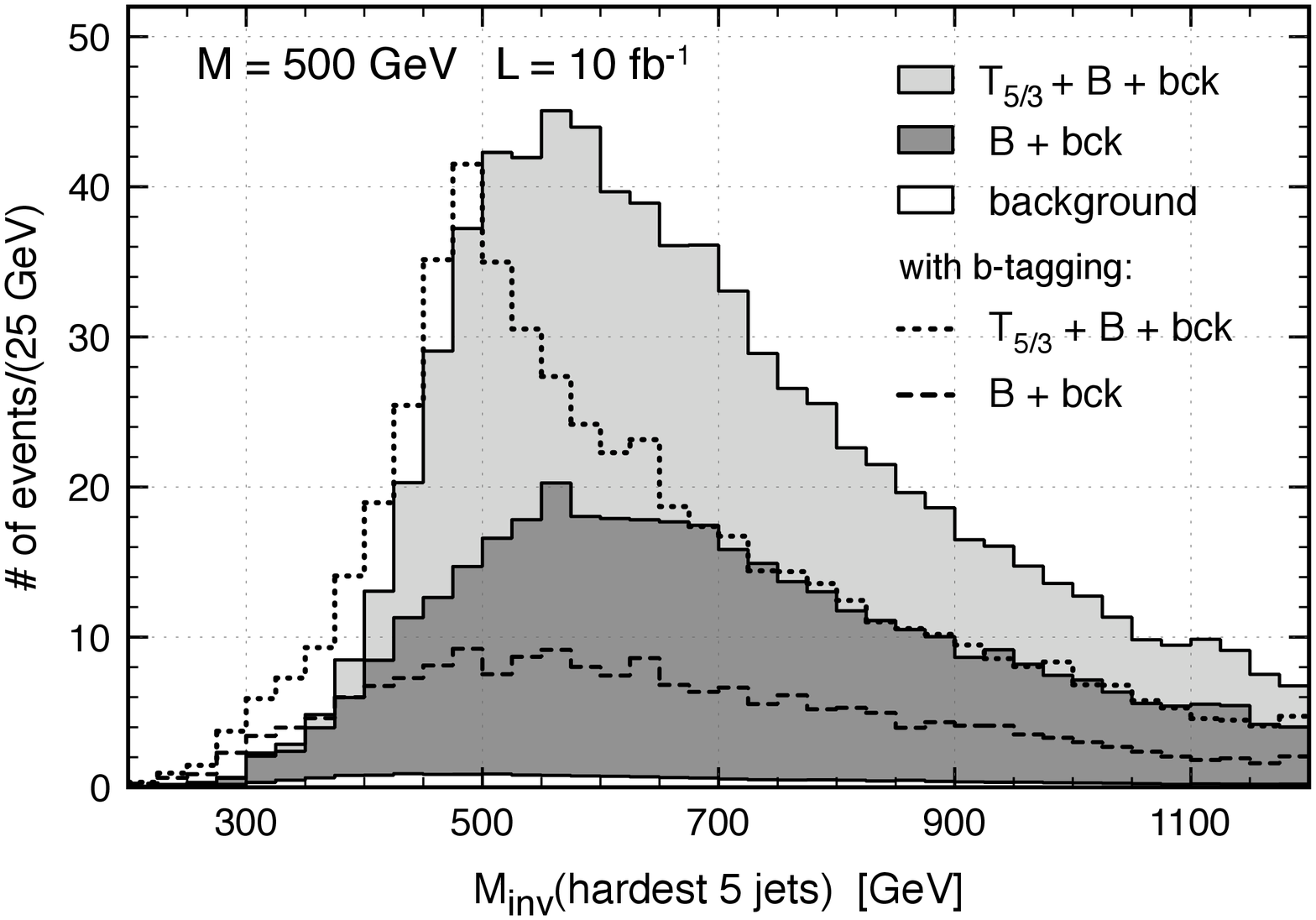} \\[0.2cm]
\includegraphics[width=8.3cm]{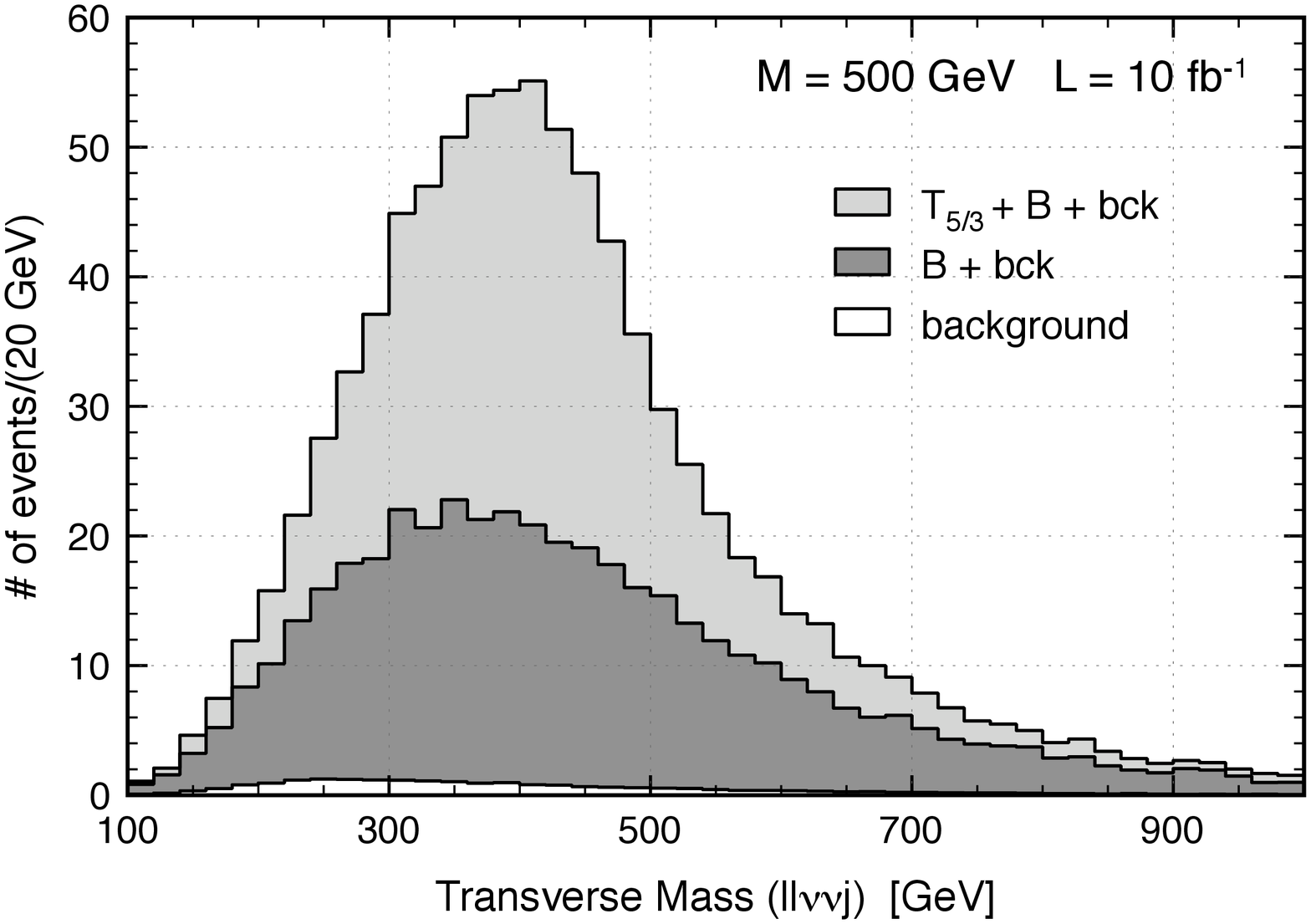}
\caption{\label{fig:TotalInvMass500} \small
Distributions after the main cuts of eq.(\ref{maincuts}) for $M=500\,\text{GeV}$: 
$a)$ Total invariant mass (upper left plot);
$b)$ Invariant mass of the hardest $5$ jets (upper right plot); 
$c)$ Transverse invariant mass of the system $(ll\nu\nu j)$, see text (lower plot).
The dotted and dashed curves in $b)$ correspond to the invariant mass of the hardest 4 jets plus 
the $b$-jet that has the largest $\Delta R$ with the softest lepton. They assume two
$b$ tags, though no $b$-tagging efficiency has been included, see text.}
\end{center}
\end{figure}

For $500\,\text{GeV}$ masses, the signal is so much larger than the background after the cuts of 
eq.(\ref{maincuts}) that a plot of the total invariant mass of the jets and the leptons
(that is: the total invariant mass of the event, excluding the missing energy) 
gives a clear and striking evidence for a resonant production at $M_{inv}(\text{tot}) = 2 M$, 
see Fig.~\ref{fig:TotalInvMass500}.
%%%%%%%%%%%%%%%%%%%%%%%%%%%%%%%%%%%%%%%%%%%%%%%%%%%
%%%%%%%%%%%%%%%%%%%%%%%%%%%%%%%%%%%%%%%%%%%%%%%%%%%
Moreover, by plotting the invariant mass of the hardest $5$ jets 
one gets the additional indication of the presence of a resonance 
at $M_{inv}(5j)\sim M$ in the scenario with $T_{5/3}$, 
see right upper plot of Fig.~\ref{fig:TotalInvMass500}.
Although this latter resonant peak is also quite evident, it is centered at values slightly
larger than the true $T_{5/3}$ mass, suggesting that the hardest 5 jets not always coincide
with those originating from the hadronic decay of the new heavy fermion.
A better resolution of the $T_{5/3}$ mass can be obtained by demanding two $b$ tags, and plotting
the invariant mass of the hardest 4 jets plus the $b$-jet that has the largest $\Delta R$ 
with the softest lepton.
This last requirement is useful to reduce the combinatorial background eliminating the $b$ from 
the semileptonic decay of the top, as the latter 
is typically very boosted and its decay products emerge in a small angular cone.
The dotted and dashed curves in the right upper plot of Fig.~\ref{fig:TotalInvMass500} show the invariant
mass distributions obtained in this way.  No $b$-tagging efficiency factor
has been included,~\footnote{The $b$-tagging algorithm that we have used, 
from the MadGraph/MadEvent distribution~\cite{MGdistribution}, has an intrinsic tagging 
efficiency that we have however rescaled out.}
in order not to commit to any specific value.
Since typical $b$-tagging efficiencies at the LHC are of the order $\eps_b \sim 0.5$,
the final distribution will be rescaled by a factor $\eps_b^2 \sim 0.25$, suggesting 
that $b$-tagging is probably not worth in the initial discovery phase, but it will be
quite effective to obtain a better mass resolution 
after having accumulated sufficient statistics.

Finally, some further crucial information on the kinematics of the events comes
from the two same-sign leptons. 
In the case of the $T_{5/3}\bar T_{5/3}$ events, 
one would like to reconstruct the leptonic decay of the second heavy fermion, 
although this is complicated by the presence of two neutrinos.
Here we consider only a very simple reconstruction procedure,
leaving more sophisticated approaches to future analyses.
Figure~\ref{fig:TotalInvMass500}, bottom plot, 
shows the transverse invariant mass of the system  
[two leptons + two neutrinos + jet closest to the softest lepton]
-- where ``closest'' here means ``with the smallest $\Delta R$'' -- defined as
\begin{equation} \label{MT}
\begin{gathered}
M_{T}^2(ll\nu\nu j) = \left( E_T(llj)+E_T(\nu\nu) \right)^2 - 
 |\vec p_T(llj) + \vec{\not\! p}_T|^2\, , \\[0.2cm]
E_T(llj) \equiv \sqrt{|\vec p_T(llj)|^2+M_{inv}(llj)^2}\, , 
\qquad E_T(\nu\nu) \equiv |\vec{\not\! p}_T|\, .
\end{gathered}
\end{equation}
In the scenario with $T_{5/3}$ partners, 
the transverse mass distribution has an
approximate edge at $M_{T}(ll\nu\nu j)\sim M$ 
due to the resonant leptonic decay, 
\footnote{
The edge is only approximate because of the omission of the unknown 
invariant mass of the system of the two neutrinos in the definition (\ref{MT}).}
while it is smoother in the other scenario with only the $B$
(where no resonance is expected in the system of the two leptons).

\begin{figure}[t!]
\begin{center}
\includegraphics[width=8.25cm]{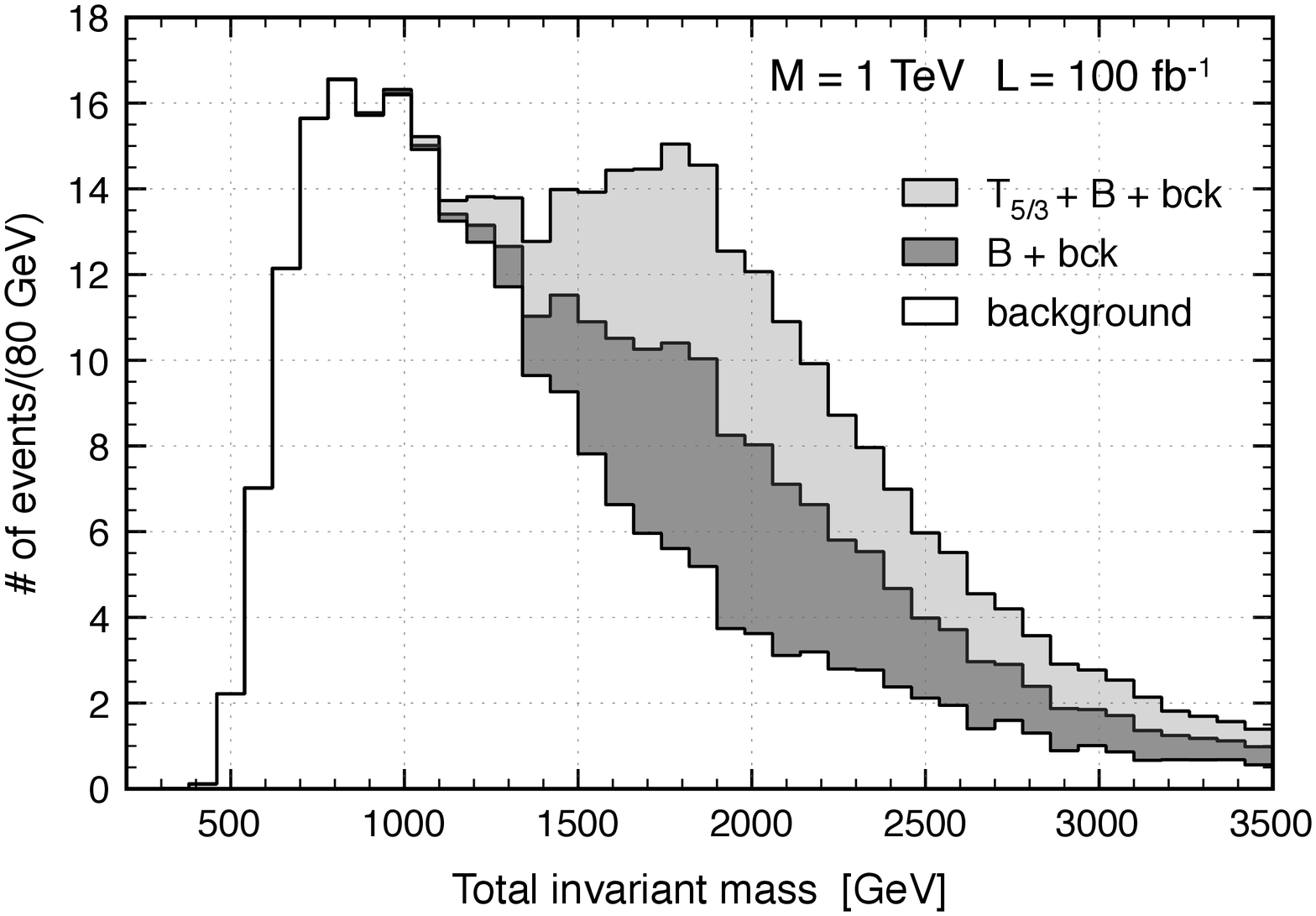} \hspace{-0.25cm}
\includegraphics[width=8.25cm]{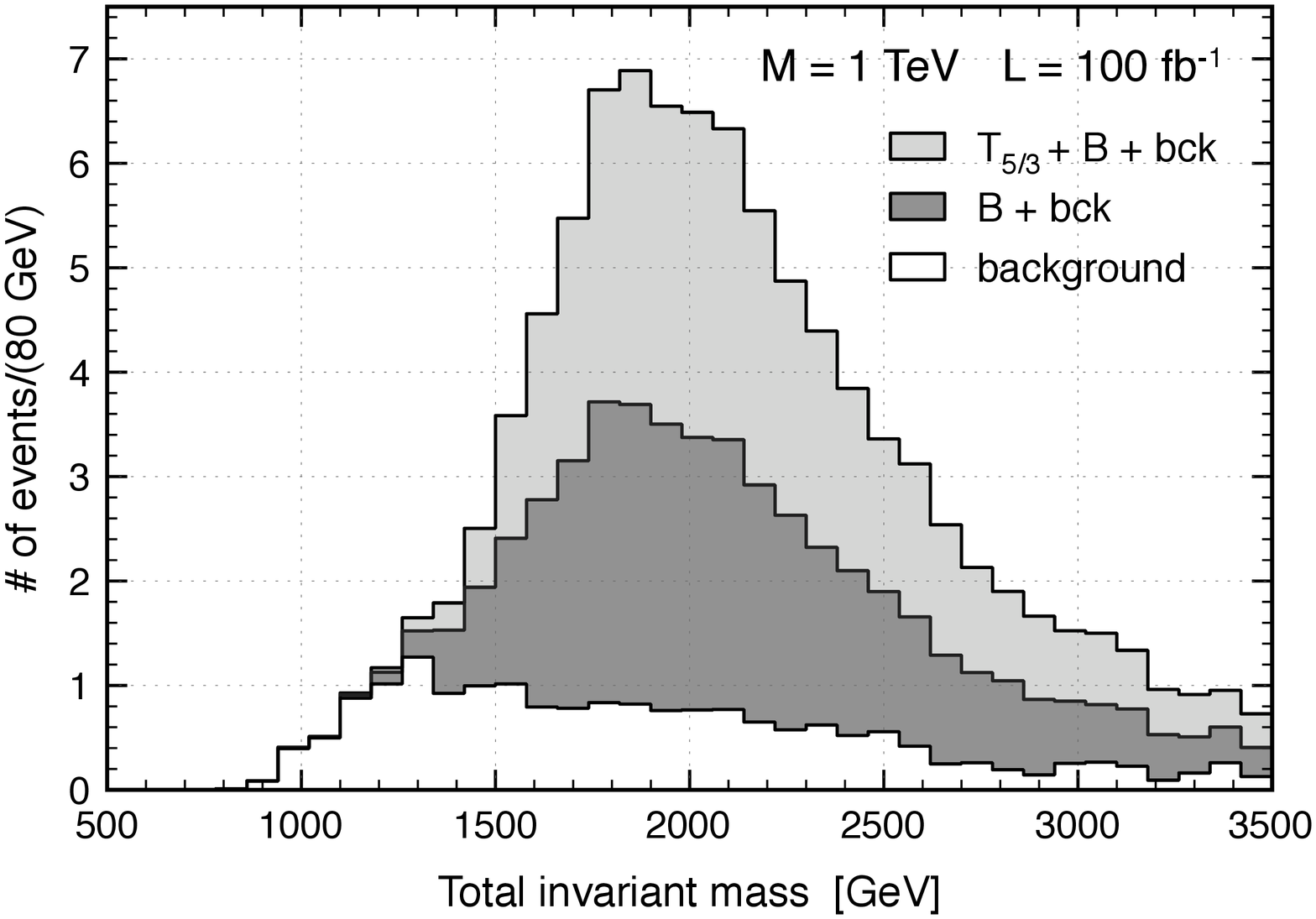} \\[0.2cm]
\includegraphics[width=8.25cm]{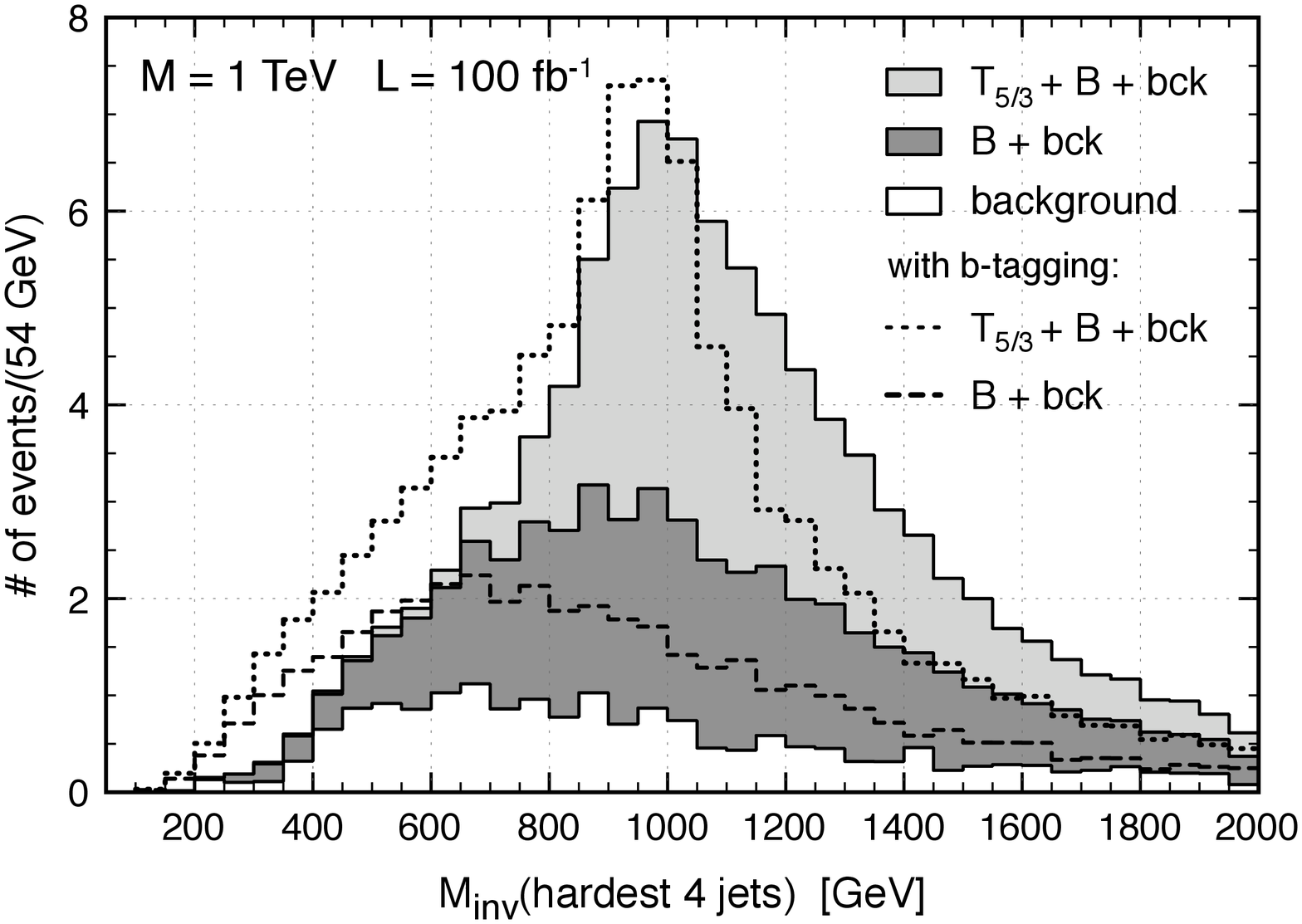} \hspace{-0.25cm}
\includegraphics[width=8.25cm]{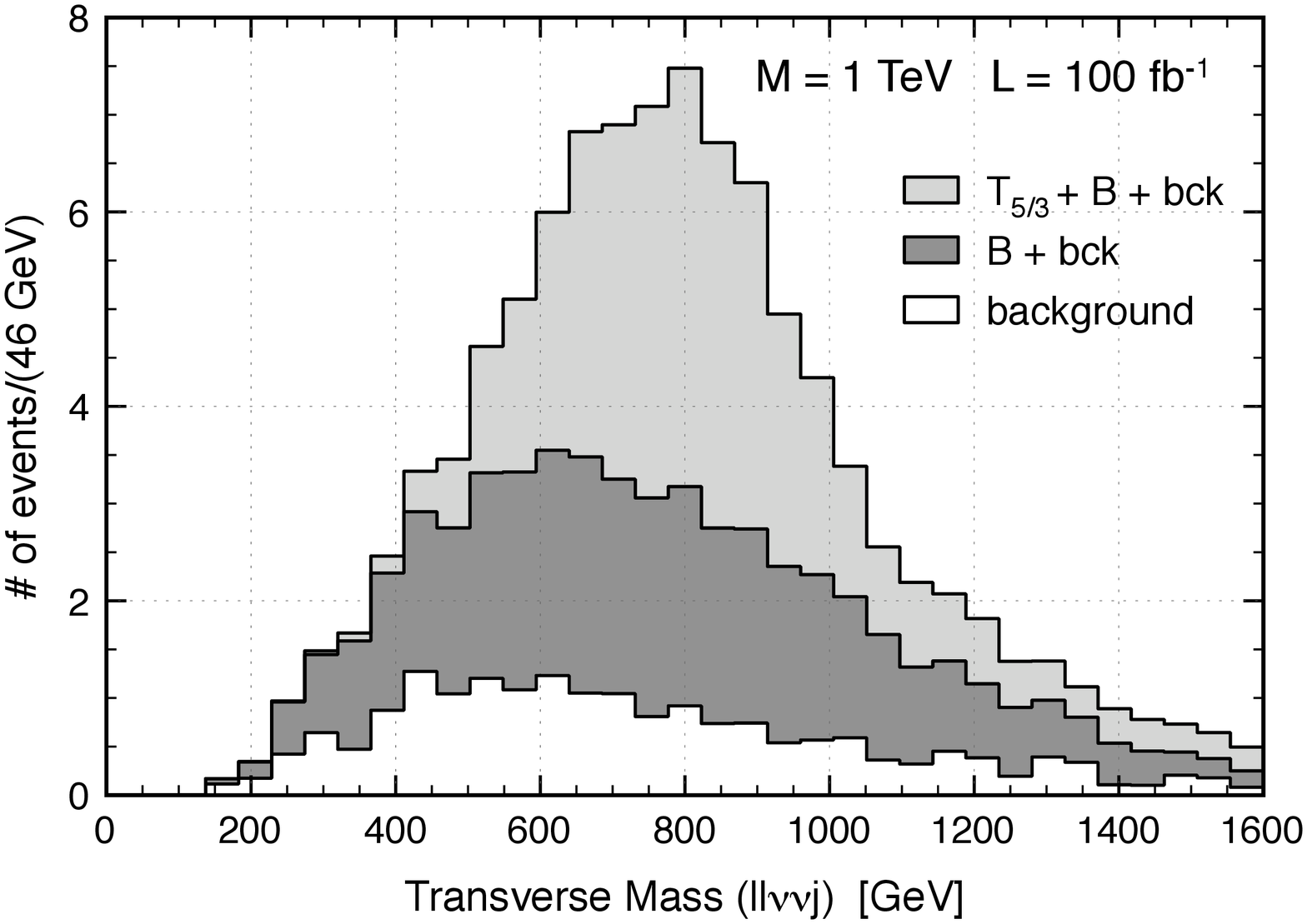}
\caption{\label{fig:TotalInvMass1000} \small
Distributions for $M=1\,\text{TeV}$: 
$a)$ Total invariant mass after the main cuts of eq.(\ref{maincuts}) (upper left plot);
$b)$ Total invariant mass after the extra cuts of eq.(\ref{discoverycuts}) (upper right plot);
$c)$ Invariant mass of the hardest $4$ jets, after the extra cuts of eq.(\ref{discoverycuts})
   (lower left plot); 
$d)$ Transverse invariant mass of the system $(ll\nu\nu j)$, after the extra cuts of 
   eq.(\ref{discoverycuts}), see text (lower plot).
The dotted and dashed curves in $c)$ correspond to the invariant mass of the hardest 3 jets plus 
the $b$-jet that has the largest $\Delta R$ with the softest lepton. They assume two
$b$ tags, though no $b$-tagging efficiency has been included, see text. 
}
\end{center}
\end{figure}

For $1\,\text{TeV}$ masses the SM background is still larger than the signal after the cuts of
eq.(\ref{maincuts}), but the resonant peak at $M_{inv}(\text{tot}) = 2M$ is already distinguishable
in the total invariant mass distribution, see the upper left plot of Fig.~\ref{fig:TotalInvMass1000}.
%%%%%%%%%%%%%%%%%%%%%%%%%
%%%%%%%%%%%%%%%%%%%%%%%%%
To further reduce the background and isolate the resonance we have performed the 
following extra ``discovery''  cuts:
\begin{equation} \label{discoverycuts}
p_T(\text{1st jet}) \ge 200\, \text{GeV}\, , \qquad \sum_{i=1,2} |\vec p_T(l_i)| \ge 300\,\text{GeV} \, .
\end{equation}
The corresponding efficiencies are reported in Table~\ref{table:epsdisc}.
%%%%%%%%%%%%%%%%%%%%%%%%%%%%%%%%%%%
\begin{table}[t]
\begin{center}
\setlength\extrarowheight{3pt}
\begin{tabular}{|l|c|c|c|c|c|}
\hline
&  signal & \multirow{2}{*}{$t\bar t W$}
& \multirow{2}{*}{$t\bar t WW$} & \multirow{2}{*}{$WWW$} & \multirow{2}{*}{$WW$} \\
& ($M=1$ TeV) & & & & \\[0.05cm]
\hline 
Efficiencies ($\eps_{disc}$)  & 0.65 & 0.091 & 0.032 & 0.16 & 0.18 \\[0.01cm]
$\sigma\, \text{[fb]} \times BR \times \eps_{main}\times \eps_{disc}$  & 0.43 & 0.12 & 0.02 & 0.02 & 0.03 \\[0.05cm]
\hline 
\end{tabular}
\caption{\label{table:epsdisc} \small
Efficiencies of the extra ``discovery'' cuts of eq.(\ref{discoverycuts}) for the case 
$M=1\,\text{TeV}$. Here signal means either $T_{5/3}\bar T_{5/3}$ or $B\bar B$ events.}
\end{center}
\end{table}
%%%%%%%%%%%%%%%%%%%%%%%%%%%%%%%%%%%
After these cuts, similarly to the $500\,\text{GeV}$ case, 
finding the correlated resonant peaks in the total invariant mass and in the 
invariant mass of the hardest $4$ jets would give strong indication
that a pair of $T_{5/3}$ has been produced with mass $M=1\,\text{TeV}$.
This could be further confirmed by the transverse mass distribution
of the $(ll\nu\nu j)$ system. 
The presence of a resonant peak only in the total invariant mass
would instead give evidence for a $B\bar B$ pair production.
All these distributions are reported in Fig.~\ref{fig:TotalInvMass1000}.
Notice that, differently from the $500\,\text{GeV}$ case, here we have plotted the invariant mass
of the hardest $4$ (not $5$) jets, since, as showed in section~\ref{sec:Strategy}, for $M=1\,\text{TeV}$ 
the signal typically contains one double jet from a boosted $W$ decay. 
Accordingly, the dotted and dashed curves in plot $c)$ of Fig.~\ref{fig:TotalInvMass1000}, obtained by 
requiring two $b$-tags, correspond to the invariant mass of the hardest 3 jets plus the $b$-jet that has the 
largest $\Delta R$ with the softest lepton.

By counting the number of signal and background events that 
pass the main cuts of eq.(\ref{maincuts}) (plus those of eq.(\ref{discoverycuts}) in the $M=1\,\text{TeV}$ case), 
one can estimate the statistical significance of the signal over 
the background, as well as the minimum integrated luminosity required for a discovery.
We define the latter to be the integrated luminosity for which a goodness-of-fit test of the SM-only hypothesis
with Poisson distribution gives a p-value = $2.85\times 10^{-7}$~\cite{p-value}.~\footnote{This p-value 
corresponds to a $5\sigma$ significance in the limit of a gaussian distribution.}
Our results are reported in Table~\ref{table:significances}.
%%%%%%%%%%%%%%%%%%%%%%%%%%%%%%%%%%%%%%%%%%%%%%%%%%%%%%%%%%%
\begin{table}[t]
\begin{center}
\setlength\extrarowheight{3pt}
\begin{tabular}{|l|l|c|c||r @{} l|}
\hline
& & ${\cal S}$ & ${\cal B}$ & \multicolumn{2}{c|}{$L_{disc}$}\\
\hline
\multirow{2}{*}{$M=500$ GeV} & $T_{5/3}+B$ & 864 & 23 & $56$ & $\,\text{pb}^{-1}$  \\
                             & $B$ only & 424 & 23 & $147$ & $\,\text{pb}^{-1}$  \\[0.05cm]
\hline
\multirow{2}{*}{$M=1$ TeV}   & $T_{5/3}+B$ & 83 & 19 & $15$ & $\,\text{fb}^{-1}$  \\
                             & $B$ only & 40 & 19 & $48$ & $\,\text{fb}^{-1}$  \\[0.05cm]
\hline
\end{tabular}
\end{center}
\caption{\label{table:significances} \small
Number of signal (${\cal S}$) and background (${\cal B}$) events
that pass the main cuts of eq.(\ref{maincuts}) (eq.(\ref{maincuts}) and eq.(\ref{discoverycuts})) 
with $L=10\,\text{fb}^{-1}$ ($L=100\,\text{fb}^{-1}$) for $M = 500\,\text{GeV}$ ($M = 1\,\text{TeV}$).
The last column reports the corresponding integrated luminosity needed for the
discovery ($L_{disc}$), 
as computed by means of a goodness-of-fit test with Poisson distribution
and p-value = $2.85\times 10^{-7}$ (see text).
}
\end{table}
%%%%%%%%%%%%%%%%%%%%%%%%%%%%%%%%%%%%%%%%%%%%%%%%%%%%%%%%%%%
In the most favorable case where both $T_{5/3}$ and $B$ partners exist and have mass $M = 500\,\text{GeV}$,
a discovery will need only $\sim 56\,\text{pb}^{-1}$. In the 1 TeV case, the theoretical uncertainty on the SM background can reduce the significance of the observed excess. Nevertheless, 
our estimates should still be conservative as
we did not include any K-factor in our
analysis, although it is known that next-to-leading order corrections enhance the signal cross section
by $\sim 80\%$ ($\sim 60\%$) for $M = 500\,\text{GeV}$ ($M = 1\,\text{TeV}$)~\cite{Bonciani:1998vc}.
Even a common K-factor $\kappa$ for both the signal and the background would imply a statistical significance
larger by a factor $\sim\sqrt{\kappa}$, as well as a discovery luminosity smaller by the same factor.
After an excess of events has been established, the compatibility with $B$ or $T_{5/3}$ pair production can be demonstrated using the shapes of the signal distributions of
 Fig.~\ref{fig:TotalInvMass500} and~\ref{fig:TotalInvMass1000} for a given value of the mass.

\section{Mass Reconstruction}
\label{subsec:reconstruction}

%%%%%%%%%%%%%%%%%%%%%%%%%%%%%%%%%%%%%%%%%%%%%%%%%%%%%%%%%%%%%%%%%%%%%%%%%%%%
\begin{figure}[t]
\begin{center}
\includegraphics[width=8.3cm]{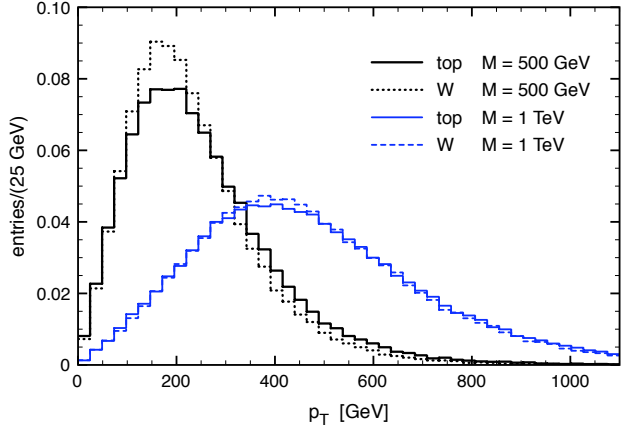}
\caption{\label{fig:pTWandtop} \small
Distributions of the $p_T$ of the $W$ and $t$ in signal events,
normalized to unit area.}
\end{center}
\end{figure}
%%%%%%%%%%%%%%%%%%%%%%%%%%%%%%%%%%%%%%%%%%%%%%%%%%%%%%%%%%%%%%%%%%%%%%%%%%%%
%%%%%%%%%%%%%%%%%%%%%%%%%%%%%%%%%%%%%%%%%%%%%%%%%%%%%%%%%%%
\begin{table}[t]
\begin{center}
\setlength\extrarowheight{3pt}
\begin{tabular}{|l|c|c|c|c|c|}
\hline
& signal & \multirow{2}{*}{$t\bar t W$}
& \multirow{2}{*}{$t\bar t WW$} & \multirow{2}{*}{$WWW$} & \multirow{2}{*}{$WW$} \\
& ($M=500$ GeV) & & & & \\
\hline 
$\eps_{2W}$    & 0.62 & 0.36 & 0.49 & 0.29 & 0.15 \\[0cm]
$\eps_{top}$   & 0.65 & 0.56 & 0.64 & 0.35 & 0.35 \\[0.05cm]
\hline 
\end{tabular}
\end{center}
\caption{\label{table:effrec500GeV} \small
Individual efficiencies for the reconstruction of two $W$'s ($\eps_{2W}$) and one top ($\eps_{top}$) 
using the algorithm  and the cuts described in the text for the case $M=500\,\text{GeV}$. The total efficiency for the top reconstruction is $\eps_{2W} \times \eps_{top}$.
}
\end{table}
%%%%%%%%%%%%%%%%%%%%%%%%%%%%%%%%%%%%%%%%%%%%%%%%%%%%%%%%%%%

More direct evidence for the production of a pair of $T_{5/3}$ or $B$ comes from
reconstructing the hadronically decayed top quark and $W$ boson, as well as 
from the distribution of 
the invariant mass of their system.

In the $M=500\,\text{GeV}$ case, we first select the events where two $W$'s can be 
simultaneously reconstructed, each $W$ candidate being formed by a pair of jets
with invariant mass in the window $|M(jj)-m_W|\leq 20\,\text{GeV}$. To avoid wrong pairings
and reduce the fake ones from the background, we impose the following cuts: 
\begin{align}
& \Delta R_{jj}\leq 1.5\, , \quad |\vec p_T(W)|\geq 100\,\text{GeV} & &
 \text{on the first $W$ candidate}\, ; \\[0.2cm]
& \Delta R_{jj}\leq 2.0\, , \quad |\vec p_T(W)|\geq 30\,\text{GeV} & &
 \text{on the second $W$ candidate}\, .
\end{align}
The $p_T$ cuts, in particular, have been optimized using the distributions of 
Fig.~\ref{fig:pTWandtop}.
If more than one pair of $W$ candidates exists which satisfies the above cuts, we select that
with the smallest $\chi^2 = \Delta R_{jj}^2(\text{1st pair})+\Delta R_{jj}^2(\text{2nd pair})$.
We then reconstruct the top by forming $Wj$ pairs, made of one $W$ and one of the remaining jets, 
with invariant mass in the window $|M(Wj)-m_t|\leq 25\,\text{GeV}$.
If more than one top candidate exists, we select that with invariant mass closest to $m_t$.
We discard events where no top can be reconstructed. 
The efficiencies of this reconstruction algorithm are reported in Table~\ref{table:effrec500GeV}.
%%%%%%%%%%%%%%%%%%%%%%%%%%%%%%%%%%%%%%%%%%%%%%%%%%%%%%%%%%%%%%%%%

\begin{figure}[t]
\begin{center}
\includegraphics[width=8.3cm]{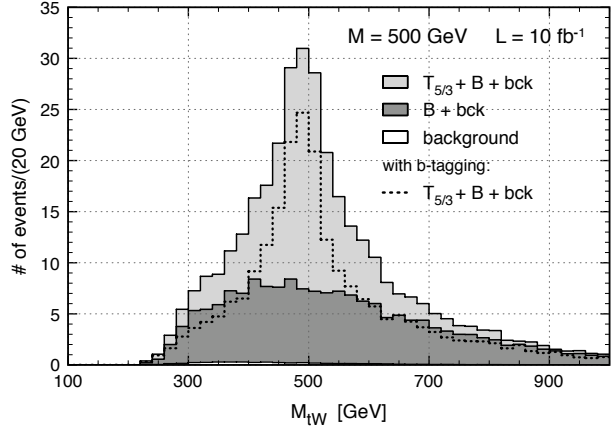}
\caption{\label{fig:Mtw_500GeV}  \small
Invariant mass of the $Wt$ system for $M=500\,\text{GeV}$ with $L=10\,\text{fb}^{-1}$.
The dotted curve refers to
the case in which $b$-tagging is performed in the reconstruction, see text.
It assumes two $b$ tags, though no $b$-tagging
efficiency has been included.}
\end{center}
\end{figure}

The distribution of the $Wt$ invariant mass is plotted in Fig.~\ref{fig:Mtw_500GeV}.
%%%%%%%%%%%%%%%%%%%%%%%%%%%%%%%%%%%%%%%%%%%%%%%%%%%%%%%%%%%%%%%%%%%%%%%%%%%%
%%%%%%%%%%%%%%%%%%%%%%%%%%%%%%%%%%%%%%%%%%%%%%%%%%%
As expected, in the scenario with $T_{5/3}$ partners there is a 
resonant peak centered at $M_{T_{5/3}}=500\,\text{GeV}$, while 
the distribution has a non-resonant, continuous shape if only $Q_{e}=-1/3$ heavy fermions exist.
The dotted curve refers to the case in which $b$-tagging is performed in the
reconstruction algorithm. In more detail, we have selected events with two $b$ tags
and we have reconstructed the top from $Wb$ pairs, excluding at the same time the $b$ jets
when selecting the $W$ jet pair candidates. As before, no $b$-tagging efficiency has been included.

In the $1\,\text{TeV}$ case the algorithm for the reconstruction of $W$ and $t$ 
has to be modified, to take into account that signal events often contain one double jet,
as shown in section~\ref{sec:Strategy}.
As the various reconstruction requirements will themselves reduce the background, we can
start our analysis imposing a set of extra cuts, in addition to those of eq.(\ref{maincuts}),
that is less aggressive than those demanded in eq.(\ref{discoverycuts}) for the discovery. 
We require:
\begin{equation} \label{reconcuts}
M_{inv}(\text{tot}) \geq 1500\,\text{GeV}\, , \quad\; 
\begin{cases} 
 p_T(\text{1st jet}) \geq 200\,\text{GeV} \\
 p_T(\text{2nd jet}) \geq 100\,\text{GeV}
\end{cases}\!\!\!\! , \quad\;
p_T(\text{1st lepton}) \geq 100\,\text{GeV}\, .
\end{equation}
The corresponding efficiencies are reported in Table~\ref{table:epsrecon}.
%%%%%%%%%%%%%%%%%%%%%%%%%%%%%%%%%%%
\begin{table}[t]
\begin{center}
\setlength\extrarowheight{3pt}
\begin{tabular}{|l|c|c|c|c|c|}
\hline
&  signal & \multirow{2}{*}{$t\bar t W$}
& \multirow{2}{*}{$t\bar t WW$} & \multirow{2}{*}{$WWW$} & \multirow{2}{*}{$WW$} \\
& ($M=1$ TeV) & & & & \\[0.05cm]
\hline 
Efficiencies ($\eps_{rec}$)  & 0.83 & 0.18 & 0.07 & 0.22 & 0.38 \\[0.01cm]
$\sigma\, \text{[fb]} \times BR \times \eps_{main}\times \eps_{rec}$  & 0.55 & 0.24 & 0.04 & 0.03 & 0.06 \\[0.05cm]
\hline 
\end{tabular}
\caption{\label{table:epsrecon} \small
Efficiencies of the extra ``reconstruction'' cuts of eq.(\ref{reconcuts}) for the case 
$M=1\,\text{TeV}$. Here signal means either $T_{5/3}\bar T_{5/3}$ or $B\bar B$ events.}
\end{center}
\end{table}
%%%%%%%%%%%%%%%%%%%%%%%%%%%%%%%%%%%
We design our strategy so as to be successful in three different situations: $i)$ no double jet is present
in the event; $ii)$ there is one double jet corresponding to the $W$ boson emitted in the
primary decay of the heavy fermion; $iii)$ there is one double jet originating from the decay of the top quark.
In the first case the reconstruction can proceed as for $M=500\,\text{GeV}$, using events with 
two reconstructed $W$'s; in the last two cases instead, the presence of one double jet implies that
only one $W$ should be required.
We thus divide the events into two samples as follows: those in which two $W$'s can be reconstructed, each made 
of a pair of jets with $|M(jj)-m_W|\leq 20\,\text{GeV}$; and those where only one $W$ can be reconstructed.
The two $W$ candidates of each event in the first sample are required to satisfy the following cuts:
\begin{align}
& \Delta R_{jj}\leq 0.7\, , \quad |\vec p_T(W)|\geq 250\,\text{GeV} & &
 \text{on the first $W$ candidate}\, ; \\[0.2cm]
& \Delta R_{jj}\leq 1.5\, , \quad |\vec p_T(W)|\geq 80\,\text{GeV} & &
 \text{on the second $W$ candidate}\, . 
\label{2ndW}
\end{align}
Events of the second sample are instead selected imposing the cuts of eq.(\ref{2ndW}) on their 
$W$ candidate. The efficiencies for the reconstruction of two and one $W$ (equal to the percentage of the
total events classified respectively in the first and second sample) are reported in Table~\ref{table:effrec1TeV}.
%%%%%%%%%%%%%%%%%%%%%%%%%%%%%%%%%%%%%%%%%%%%%%%%%%%%%%%%%%%
\begin{table}[t]
\begin{center}
\setlength\extrarowheight{5pt}
\begin{tabular}{|l|c|c|c|c|c|}
\hline
&  signal & \multirow{2}{*}{$t\bar t W$}
            & \multirow{2}{*}{$t\bar t WW$} & \multirow{2}{*}{$WWW$} & \multirow{2}{*}{$WW$} \\[-0.1cm]
& $(M=1\,\text{TeV})$ & & & & \\[0.1cm]
\hline  
$\eps_{2W}$   & 0.31 & 0.15 & 0.23 & 0.16 & 0.071 \\[-0.1cm]
$\eps_{1W}$   & 0.57 & 0.62 & 0.59 & 0.58 & 0.49 \\[0.1cm]
\hline 
$\eps_{top}^{[2W]}(t=Wj)$  & 0.62 & 0.56 & 0.62 & 0.11 & 0.13 \\[0.08cm]
$\eps_{top}^{[1W]}(t=Wj)$  & 0.44 & 0.56 & 0.53 & 0.22 & 0.20 \\[0.08cm]
$\eps_{top}(t=jj)$       & 0.18 & 0.04 & 0.06 & 0.06 & 0.07 \\[0.1cm]
\hline 
\end{tabular}
\end{center}
\caption{\label{table:effrec1TeV} \small
Efficiencies of the algorithms and the cuts described in the text for the case $M=1\,\text{TeV}$:
reconstruction of two ($\eps_{2W}$) or one ($\eps_{1W}$) $W$'s; reconstruction of the
top as $Wj$ using events with two ($\eps_{top}^{[2W]}(t=Wj)$) or one ($\eps_{top}^{[1W]}(t=Wj)$) $W$'s; 
reconstruction of the top from a pair of jets ($\eps_{top}(t=jj)$).
}
\end{table}
%%%%%%%%%%%%%%%%%%%%%%%%%%%%%%%%%%%%%%%%%%%%%%%%%%%%%%%%%%%

\begin{figure}[t]
\begin{center}
\includegraphics[width=8.25cm]{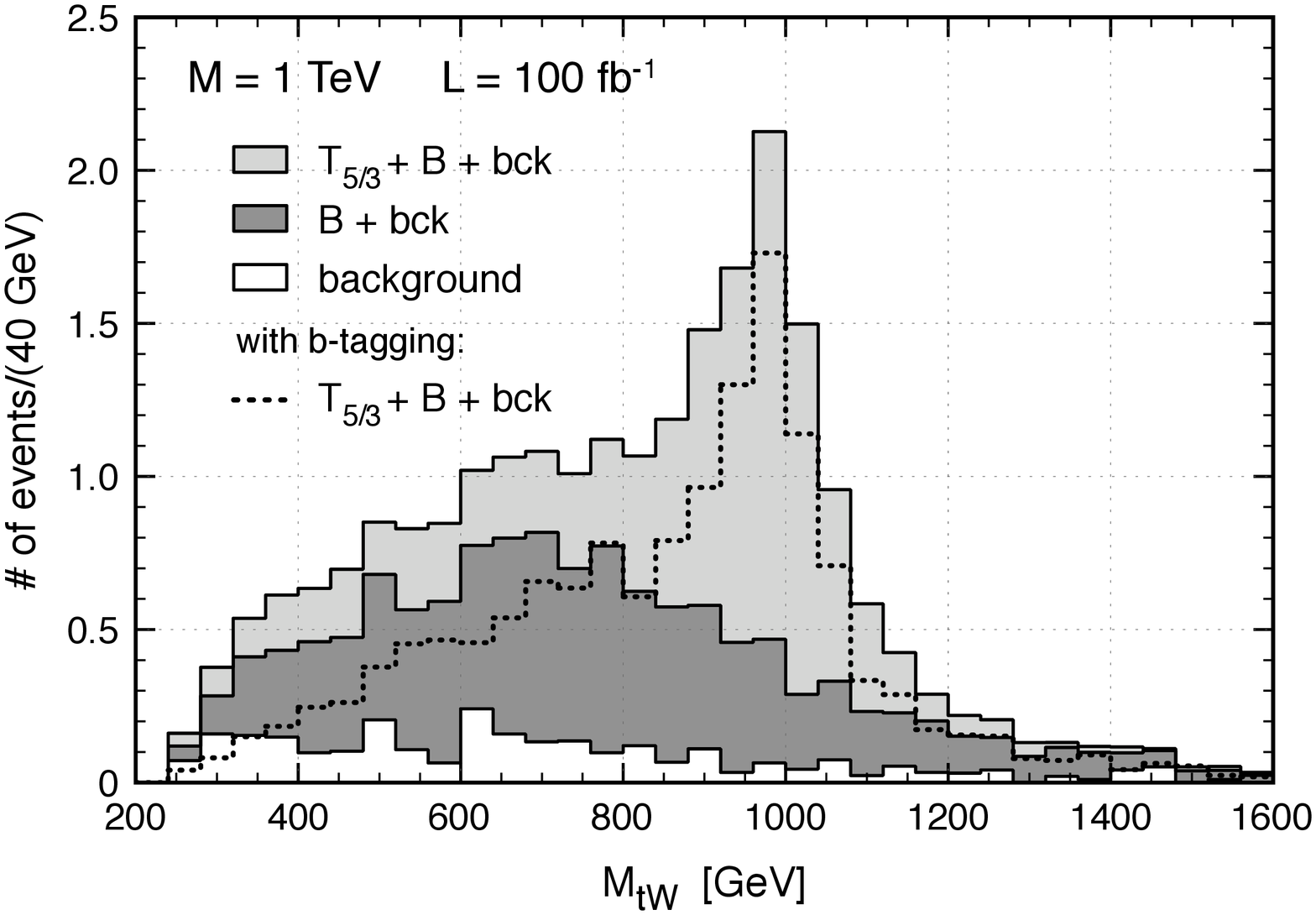} \hspace{-0.28cm}
\includegraphics[width=8.25cm]{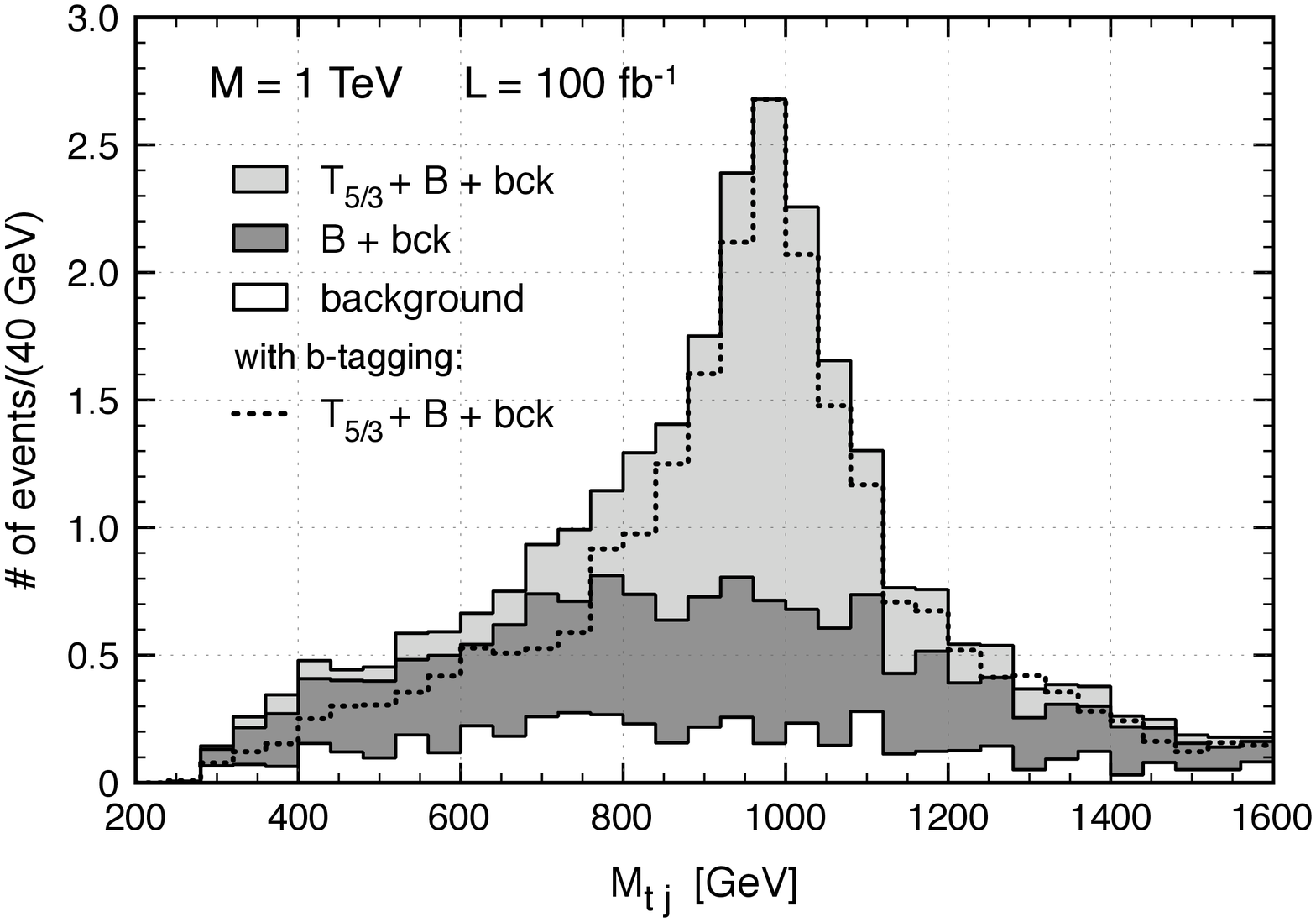} \\[0.2cm]
\includegraphics[width=8.25cm]{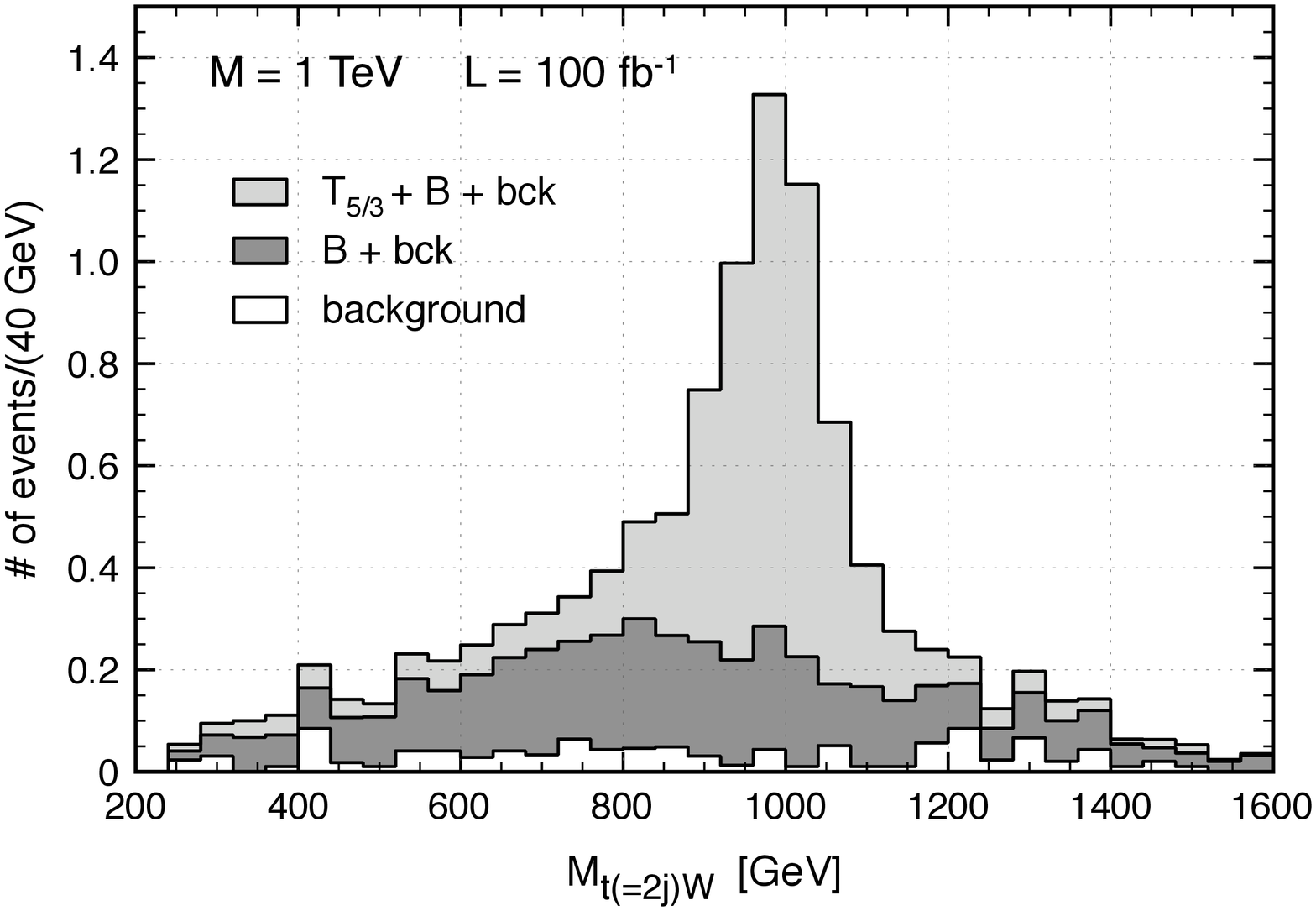}  \hspace{-0.28cm}
\includegraphics[width=8.25cm]{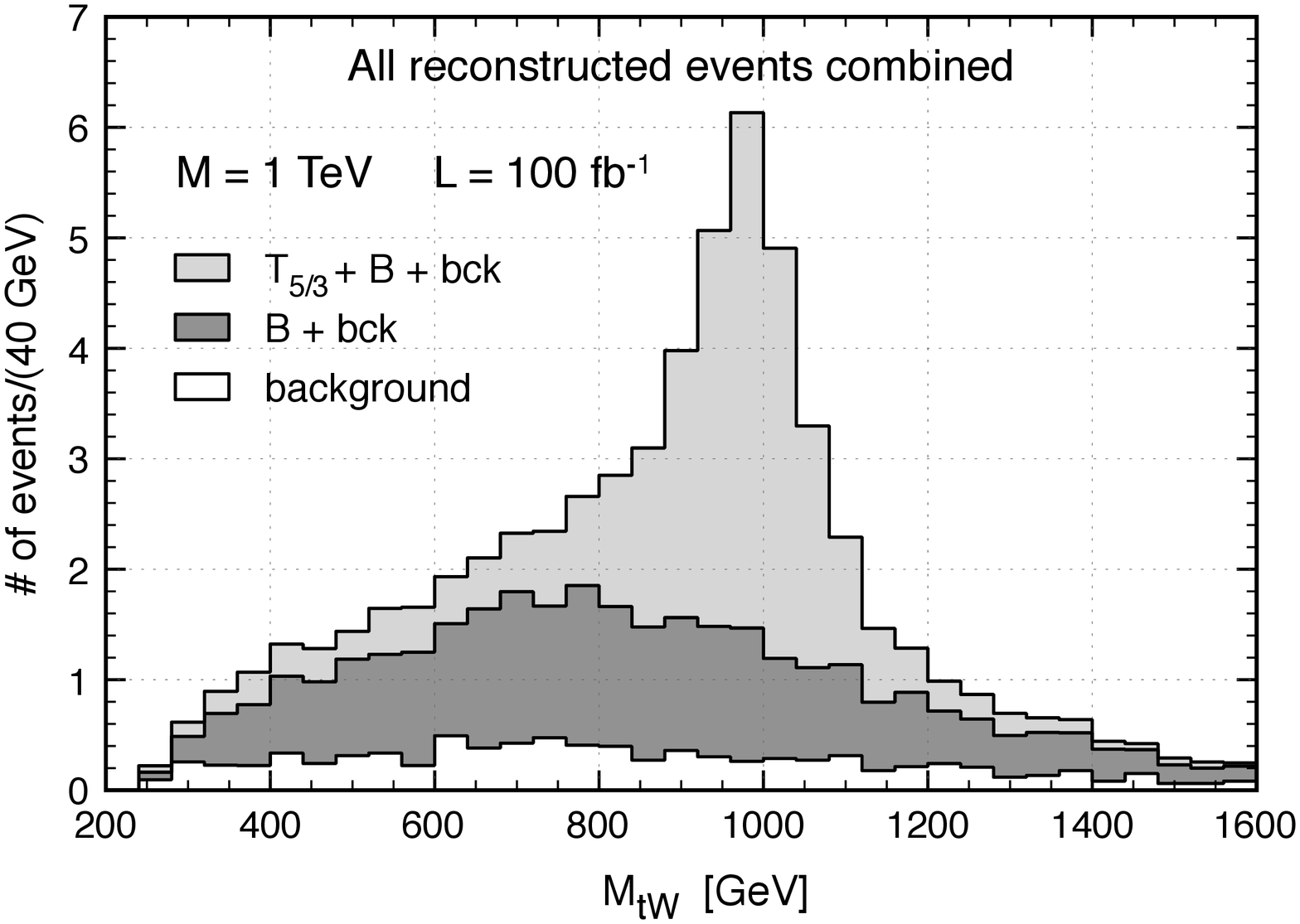}
\caption{\label{fig:Mtw_1TeV}  \small
Invariant mass of the $Wt$ system, for $M=1\,\text{TeV}$ and $L=100\,\text{fb}^{-1}$,
obtained following the three reconstruction procedures described in the text (first
three plots).
The dotted curves in the first and second plot show the effect of performing the
$b$-tagging in the top reconstruction. They assume two $b$ tags, though no $b$-tagging
efficiency has been included.
The lower right plot shows the total $Wt$ distribution obtained by combining the events
of the first three plots.}
\end{center}
\end{figure}

We thus reconstruct one top from $Wj$ pairs with $|M(Wj)-m_t|\leq 25\,\text{GeV}$, 
as for $M=500\,\text{GeV}$, using events of the first 
sample (those with two $W$'s). Events where no top can be reconstructed are removed. 
The corresponding efficiency, labeled as $\eps_{top}^{[2W]}(t=Wj)$, is reported in Table~\ref{table:effrec1TeV}.
The final invariant mass of the $Wt$ pair is plotted in Fig.~\ref{fig:Mtw_1TeV}, upper left plot.
%%%%%%%%%%%%%%%%%%%%%%%%%%%%%%%%%%%%%%%%%%%%%%%%%%%%%%%%%%%%%%%%%%%%%%%%%%%%
%%%%%%%%%%%%%%%%%%%%%%%%%%%%%%%%%%%%%%%%%%%%%%%%%%%%%
The dotted curve refers to the case in which $b$-tagging has been performed in the reconstruction,
according to the same procedure adopted in the $M=500\,\text{GeV}$ case.

Events from the second sample (those with one $W$) are used to reconstruct the top in two possible ways:
first, the reconstruction is attempted selecting $Wj$ pairs with $|M(Wj)-m_t|\leq 25\,\text{GeV}$; 
in case of unsuccess, we then try to reconstruct the top forming pairs of jets with
$|M(jj)-m_t|\leq 25\,\text{GeV}$, where one of the two is assumed to be a double jet.
If more than one top candidate exists, we select that with invariant mass closest to $m_t$.
If instead none is found the event is removed. The efficiencies of these two reconstruction procedures,
respectively labeled as $\eps_{top}^{[1W]}(t=Wj)$ and $\eps_{top}(t=jj)$,
are reported in Table~\ref{table:effrec1TeV}.
Events with one $W$ in which the top is reconstructed as $Wj$ are those where the double jet corresponds to 
the $W$ boson emitted in the heavy fermion primary decay. It turns out that such double jet is usually
the hardest among the remaining jets (those not involved in the top reconstruction).
We thus plot the invariant mass of $tj$, where $j$ is the hardest other jet with 
$p_T\geq 80\,\text{GeV}$, see the upper right plot of Fig.~\ref{fig:Mtw_1TeV}.
As before, the dotted curve shows the effect of performing the $b$-tagging in the top reconstruction.
For events with one $W$ and the top reconstructed from a pair of jets we plot instead the $tW$ invariant mass,
see the lower left plot of Fig.~\ref{fig:Mtw_1TeV}.

As illustrated in the first three plots of Fig.~\ref{fig:Mtw_1TeV}, all three different methods that
we have described for reconstructing the hadronically decayed $T_{5/3}$ are quite successful: the resonant
peak at $M_{T_{5/3}}=1\,\text{TeV}$ is always clearly distinguishable over the non-resonant distribution due to the $B$.
The presence of the peak would thus give the ultimate evidence in favor of the $T_{5/3}$, whereas its absence
would rather indicate that a pair of $B$'s has been produced.
To enhance the statistical significance, the events from the three plots can be combined into a single
distribution, shown in the lower right plot of Fig.~\ref{fig:Mtw_1TeV}.

\section{Discussion and Outlook}

The results of sections~\ref{subsec:discovery} and~\ref{subsec:reconstruction} show
that the analysis of final states with two same-sign leptons  at the LHC
is an extremely promising
method to discover the top partners $B$ and $T_{5/3}$.
By requiring two same-sign leptons one avoids the large $t\bar t$ background and selects
a particularly clean channel where evidence for 
the existence of the heavy fermions
could come in the early phase of the LHC.
The estimate of section~\ref{subsec:discovery} suggests that a  discovery could 
be claimed already with $\sim 50\,\text{pb}^{-1}$ or $\sim 150\,\text{pb}^{-1}$ for $M=500$ GeV, respectively 
if both $B$ and $T_{5/3}$ or only $B$ exist.
Even without $b$-tagging, and before reconstructing the hadronically decayed $W$ and top,
one can have a first crucial indication on the value of the mass of the 
heavy fermions from the distributions of the total invariant mass and the 
invariant mass of the hardest $5$ or $4$ jets.
The presence of a resonant peak in both distributions, respectively at $M_{inv}(\text{tot})\sim 2M$ 
and $M_{inv}(\text{hardest 5 or 4 } jets)\sim M$, 
would be specific evidence for the production of the $T_{5/3}$.

Although the use of $b$-tagging can increase the resolution of the resonant peaks, hence their
statistical significance, the ultimate evidence for the discovery of $T_{5/3}$ would come
from its reconstruction in the $Wt$ invariant mass. 
As our explorative study indicates, the strategy to follow in that case will need to be optimized according
to the value of the heavy fermion mass $M$. 
In general, it will be preferable to suitably choose and tune the jet algorithm to individually resolve 
as many jets as possible. In the case of a cone algorithm, this means choosing a not too
large cone size. We found that $\Delta R=0.4$ gives good results, as it permits to resolve
all the jets from the decay of the top and the $W$ for $M=500\,\text{GeV}$, while only one
double jet is typically present in the signal events for $M=1\,\text{TeV}$.
In this respect our analysis differs from that of Ref.~\cite{Skiba:2007fw}, 
where the proposed strategy to reconstruct $1\,\text{TeV}$ heavy bottoms was
that of selecting and pairing jets with invariant mass close to $m_t$ and $m_W$.
In the case of the $B$, its full reconstruction will be only possible
by analyzing events with one or two opposite-sign leptons.
For that purpose, the first rough indication on the value of $M$ extracted from the same-sign dilepton
events will serve to guide the analysis and optimize the cuts needed to kill the large SM background.
In this sense, the use of all final states with different lepton multiplicities will permit
to discriminate different scenarios where only one or both top partners exist.

Ultimately,
a crucial information to understand the origin and the role of the heavy fermions would
come from the measurement of their decay width, which will in turn lead to a determination of
their couplings $\lambda_{T_{5/3},B}$.
As already stressed before, a large value of $\lambda_{T_{5/3},B}$
will be strong circumstantial evidence for the compositeness of the heavy fermions.
Extracting the decay width from the invariant mass distribution will be challenging,
as one will have to cope with the issue of jet energy resolution.
Most likely, a measurement will be possible only with large statistics and will require
more sophisticated $W$ and $t$ reconstruction techniques.

In this analysis, we only considered the model-independent pair production of the top partners, neglecting their
single production. At the LHC the latter proceeds through $Wt$ fusion, via the diagram of 
Fig.~\ref{fig:singleproduction}, and leads to final states with $t\bar t W+jets$. 
It will thus contribute to the same-sign dilepton channel,
enhancing the significance of the new physics signal over the SM background.
Its effect will be more important
for heavier masses $M$ (for which the pair-production cross section is more suppressed) 
and larger couplings $\lambda_{T_{5/3},B}$.
However, due to the absence of a second, hadronically decaying top partner, 
events from single production will not give a resonant contribution
to the $Wt$ invariant mass distribution, or to the invariant mass of the hardest $4$ or $5$ jets.
In this sense, the inclusion of single production should not dramatically affect the results
of our simplified analysis, and it could even lead to a larger statistical significance in the first 
discovery phase.
It is clear, however, that a dedicated analysis will be required to assess the actual
importance of single production, and to determine its potentialities in extending the LHC discovery 
reach for larger values of the heavy masses~$M$.

Given the strong theoretical motivations for a search of the heavy partners of the top,
we think that our explorative study would also deserve to be followed by a dedicated
experimental investigation. Our results suggest that the same-sign dilepton channel might 
be one of the golden modes to discover the top partners $B$ and $T_{5/3}$, but only a complete
analysis with a full simulation of the detector effects,
an exact calculation of the $Wl^+ l^-+jets$ and $t\bar t+jets$ backgrounds,
and the use of fully realistic reconstruction techniques
will eventually establish its ultimate potentialities.

\section*{Acknowledgments}

We thank A. Polosa for collaboration in the early stages of this project and for many useful discussions.
We are indebted to M. Mangano and F. Maltoni for illuminating discussions and suggestions, 
and with J. Alwall, M. Herquet and all the CP3 team for their constant and absolutely crucial
assistance on Madgraph. It is also a pleasure to thank J. A. Aguilar-Saavedra,
D. Berge, G.~Brooijmans, P.~Ciafaloni, D. Del Re, S.~Frixione, Y.~Gershtein,
M.~Moretti,  M.~Narain, F. Piccinini, M.~Pierini, R.~Rattazzi, M. Spiropulu, and J.~Tseng
for several interesting and stimulating discussions and comments.

%%%%%%%%%%%%%%%%%%%%
%% References
%%%%%%%%%%%%%%%%%%%%

\end{document}